\documentclass[aps,prd,twocolumn,superscriptaddress,showpacs,preprintnumbers,amsmath,amssymb]{revtex4-1}
\usepackage{graphicx} 
\usepackage{dcolumn}  
\usepackage{rotating}
\graphicspath{{ps}}

\begin{document}


\preprint{\vbox{ 
						\hbox{Belle Preprint {\it 2015-12}}
						\hbox{KEK Preprint {\it 2015-21}}
}}

\title{ \quad\\[1.0cm]Inclusive cross sections for pairs of identified light charged hadrons and for single protons in $e^+e^-$ at $\sqrt{s}=$ 10.58 GeV }

\noaffiliation
\affiliation{University of the Basque Country UPV/EHU, 48080 Bilbao}
\affiliation{University of Bonn, 53115 Bonn}
\affiliation{Budker Institute of Nuclear Physics SB RAS, Novosibirsk 630090}
\affiliation{Faculty of Mathematics and Physics, Charles University, 121 16 Prague}
\affiliation{Chonnam National University, Kwangju 660-701}
\affiliation{University of Cincinnati, Cincinnati, Ohio 45221}
\affiliation{Deutsches Elektronen--Synchrotron, 22607 Hamburg}
\affiliation{University of Florida, Gainesville, Florida 32611}
\affiliation{Justus-Liebig-Universit\"at Gie\ss{}en, 35392 Gie\ss{}en}
\affiliation{Gifu University, Gifu 501-1193}
\affiliation{SOKENDAI (The Graduate University for Advanced Studies), Hayama 240-0193}
\affiliation{Hanyang University, Seoul 133-791}
\affiliation{University of Hawaii, Honolulu, Hawaii 96822}
\affiliation{High Energy Accelerator Research Organization (KEK), Tsukuba 305-0801}
\affiliation{IKERBASQUE, Basque Foundation for Science, 48013 Bilbao}
\affiliation{University of Illinois at Urbana-Champaign, Urbana, Illinois 61801}
\affiliation{Indian Institute of Technology Bhubaneswar, Satya Nagar 751007}
\affiliation{Indian Institute of Technology Guwahati, Assam 781039}
\affiliation{Indian Institute of Technology Madras, Chennai 600036}
\affiliation{Indiana University, Bloomington, Indiana 47408}
\affiliation{Institute of High Energy Physics, Chinese Academy of Sciences, Beijing 100049}
\affiliation{Institute of High Energy Physics, Vienna 1050}
\affiliation{Institute for High Energy Physics, Protvino 142281}
\affiliation{INFN - Sezione di Torino, 10125 Torino}
\affiliation{Institute for Theoretical and Experimental Physics, Moscow 117218}
\affiliation{J. Stefan Institute, 1000 Ljubljana}
\affiliation{Kanagawa University, Yokohama 221-8686}
\affiliation{Institut f\"ur Experimentelle Kernphysik, Karlsruher Institut f\"ur Technologie, 76131 Karlsruhe}
\affiliation{Kennesaw State University, Kennesaw GA 30144}
\affiliation{King Abdulaziz City for Science and Technology, Riyadh 11442}
\affiliation{Department of Physics, Faculty of Science, King Abdulaziz University, Jeddah 21589}
\affiliation{Korea Institute of Science and Technology Information, Daejeon 305-806}
\affiliation{Korea University, Seoul 136-713}
\affiliation{Kyoto University, Kyoto 606-8502}
\affiliation{Kyungpook National University, Daegu 702-701}
\affiliation{\'Ecole Polytechnique F\'ed\'erale de Lausanne (EPFL), Lausanne 1015}
\affiliation{Faculty of Mathematics and Physics, University of Ljubljana, 1000 Ljubljana}
\affiliation{Luther College, Decorah, Iowa 52101}
\affiliation{University of Maribor, 2000 Maribor}
\affiliation{Max-Planck-Institut f\"ur Physik, 80805 M\"unchen}
\affiliation{School of Physics, University of Melbourne, Victoria 3010}
\affiliation{Moscow Physical Engineering Institute, Moscow 115409}
\affiliation{Moscow Institute of Physics and Technology, Moscow Region 141700}
\affiliation{Graduate School of Science, Nagoya University, Nagoya 464-8602}
\affiliation{Kobayashi-Maskawa Institute, Nagoya University, Nagoya 464-8602}
\affiliation{Nara Women's University, Nara 630-8506}
\affiliation{National Central University, Chung-li 32054}
\affiliation{National United University, Miao Li 36003}
\affiliation{Department of Physics, National Taiwan University, Taipei 10617}
\affiliation{H. Niewodniczanski Institute of Nuclear Physics, Krakow 31-342}
\affiliation{Niigata University, Niigata 950-2181}
\affiliation{Novosibirsk State University, Novosibirsk 630090}
\affiliation{Osaka City University, Osaka 558-8585}
\affiliation{Pacific Northwest National Laboratory, Richland, Washington 99352}
\affiliation{Peking University, Beijing 100871}
\affiliation{University of Pittsburgh, Pittsburgh, Pennsylvania 15260}
\affiliation{RIKEN BNL Research Center, Upton, New York 11973}
\affiliation{University of Science and Technology of China, Hefei 230026}
\affiliation{Seoul National University, Seoul 151-742}
\affiliation{Soongsil University, Seoul 156-743}
\affiliation{University of South Carolina, Columbia, South Carolina 29208}
\affiliation{Sungkyunkwan University, Suwon 440-746}
\affiliation{School of Physics, University of Sydney, NSW 2006}
\affiliation{Department of Physics, Faculty of Science, University of Tabuk, Tabuk 71451}
\affiliation{Tata Institute of Fundamental Research, Mumbai 400005}
\affiliation{Excellence Cluster Universe, Technische Universit\"at M\"unchen, 85748 Garching}
\affiliation{Toho University, Funabashi 274-8510}
\affiliation{Tohoku University, Sendai 980-8578}
\affiliation{Earthquake Research Institute, University of Tokyo, Tokyo 113-0032}
\affiliation{Department of Physics, University of Tokyo, Tokyo 113-0033}
\affiliation{Tokyo Institute of Technology, Tokyo 152-8550}
\affiliation{Tokyo Metropolitan University, Tokyo 192-0397}
\affiliation{University of Torino, 10124 Torino}
\affiliation{Utkal University, Bhubaneswar 751004}
\affiliation{CNP, Virginia Polytechnic Institute and State University, Blacksburg, Virginia 24061}
\affiliation{Wayne State University, Detroit, Michigan 48202}
\affiliation{Yamagata University, Yamagata 990-8560}
\affiliation{Yonsei University, Seoul 120-749}
  \author{R.~Seidl}\affiliation{RIKEN BNL Research Center, Upton, New York 11973} 
  \author{A.~Abdesselam}\affiliation{Department of Physics, Faculty of Science, University of Tabuk, Tabuk 71451} 
  \author{I.~Adachi}\affiliation{High Energy Accelerator Research Organization (KEK), Tsukuba 305-0801}\affiliation{SOKENDAI (The Graduate University for Advanced Studies), Hayama 240-0193} 
  \author{H.~Aihara}\affiliation{Department of Physics, University of Tokyo, Tokyo 113-0033} 
  \author{S.~Al~Said}\affiliation{Department of Physics, Faculty of Science, University of Tabuk, Tabuk 71451}\affiliation{Department of Physics, Faculty of Science, King Abdulaziz University, Jeddah 21589} 
  \author{D.~M.~Asner}\affiliation{Pacific Northwest National Laboratory, Richland, Washington 99352} 
  \author{T.~Aushev}\affiliation{Moscow Institute of Physics and Technology, Moscow Region 141700}\affiliation{Institute for Theoretical and Experimental Physics, Moscow 117218} 
  \author{R.~Ayad}\affiliation{Department of Physics, Faculty of Science, University of Tabuk, Tabuk 71451} 
  \author{V.~Babu}\affiliation{Tata Institute of Fundamental Research, Mumbai 400005} 
  \author{I.~Badhrees}\affiliation{Department of Physics, Faculty of Science, University of Tabuk, Tabuk 71451}\affiliation{King Abdulaziz City for Science and Technology, Riyadh 11442} 
% ...
\collaboration{The Belle Collaboration}

\noaffiliation
\begin{abstract}
We report the first double differential cross sections of two charged pions and kaons ($e^+e^- \rightarrow hhX$) in electron-positron annihilation as a function of the fractional energies of the two hadrons for any charge and hadron combination. The dependence of these di-hadron cross sections on the topology (same, opposite-hemisphere or anywhere) is also studied with the help of the event shape variable thrust and its axis. The ratios of these di-hadron cross sections for different charges and hadron combinations directly shed light on the contributing fragmentation functions. For example, we find that the ratio of same-sign pion pairs over opposite-sign pion pairs drops toward higher fractional energies where disfavored fragmentation is expected to be suppressed. 
These di-hadron results are obtained from a $655\,{\rm fb}^{-1}$ data sample collected near the $\Upsilon(4S)$ resonance
with the Belle detector at the KEKB asymmetric-energy $e^+ e^-$
collider.
Extending the previously published single-pion and single-kaon cross sections, single-proton ($e^+e^- \rightarrow pX$) cross sections are extracted from a $159\,{\rm fb}^{-1}$ data sub-sample.
\end{abstract}

\pacs{13.66.Bc,13.87.Fh,13.88.+e,14.20.Dh}

\maketitle

\tighten

{\renewcommand{\thefootnote}{\fnsymbol{footnote}}}
\setcounter{footnote}{0}

Quantum chromodynamics (QCD) is generally accepted as the theory of the strong interaction. It describes successfully many high-energy processes where the strong coupling is small and a perturbative treatment is applicable. However, the non-perturbative region of hadronic bound states such as the nucleon or the transition from high-energetic partons into confined hadrons cannot be described quantitatively so far. Despite the recent progress in lattice QCD, both parton distribution and fragmentation functions (FF) remain quantities that need to be obtained experimentally. 

Fragmentation functions describe the density of final-state hadrons $h$ from an initial parton $q$ with a fractional energy $z=E_h/E_q$ at a certain energy scale $Q$. Single-hadron, unpolarized fragmentation is described by the function $D_{1,q}^h(z,Q)$. The first $z$ moment, summed over all final states, corresponds to energy conservation in the transition of the initial-state parton into the total final state. 
Fragmentation functions cannot be directly accessed but can be related to observable quantities whenever hadrons appear in the final state. The most prominent connection to fragmentation functions can be found in single-hadron inclusive cross sections in electron-positron annihilation. This process provides very clean access as there are no hadrons in the initial state. These cross sections can then be related at leading order in the strong coupling, $\alpha_S$, to fragmentation functions via 
\begin{equation}
\frac{d\sigma(e^+e^-\rightarrow hX)}{dz} \propto \sum_q e_q^2 \left( D_{1,q} ^h(z,Q^2) + D_{1,\overline{q}} ^h(z,Q^2)\right),\label{eq:eq1}
\end{equation}
where the scale $Q=\sqrt{s}$ is given by the center-of-mass (CMS) energy. Many such measurements have been performed for light hadrons at a range of CMS energies at the B factories \cite{martin,babar} at LEP and SLC \cite{Buskulic:1994ft,Akers:1994ez,Abreu:1995cu,Abe:1998zs} and other facilities \cite{Brandelik:1980iy,Furmanski:1981cw,Bartel:1981sw,Althoff:1982dh,Aihara:1983ic,Schellman:1984yz,Derrick:1985wd,Aihara:1988su,Braunschweig:1988hv,Itoh:1994kb}. The different energy scales can be related via DGLAP evolution \cite{dglap}. \par
Several global analyses of the $e^+e^-$ fragmentation data have been performed \cite{kkp,akk,hkns} that provide reasonable precision on the sum of quark and anti-quark fragmentation functions. The precise Belle and BaBar data allowed to reduce the uncertainties on the gluon fragmentation function to pions, which only enters at the next-to-leading order in $\alpha_S$. However, analysis of the single-hadron cross sections from $e^+e^-$ data alone cannot distinguish nor flavor-separate quark and anti-quark fragmentation and, in particular, favored \textit{vs.~}disfavored fragmentation. Favored (disfavored) fragmentation describes the fragmentation of a parton into a hadron with (without) that parton flavor as valence content, such as $u\rightarrow \pi^+$ ($u\rightarrow \pi^-$). As the parton distribution functions are generally well known for up- and down-type flavors, semi-inclusive deep-inelastic scattering (SIDIS) \cite{hermes,compass} and hadron-collision \cite{phenixpi0,starpi0,starpik,phenixpik,alice} results provide leverage in truly global fits to obtain some flavor and charge separated information on pion and kaon fragmentation functions \cite{dss}. \par
When selecting hadron pairs in $e^+e^-$ annihilation, the cross section can be expressed at leading order in $\alpha_S$ in terms of products of fragmentation functions \cite{schweitzer}:
\begin{eqnarray}
\frac{d^2\sigma(e^+e^-\rightarrow h_1h_2X)}{dz_1dz_2} \propto&&\nonumber \\\sum_q e_q^2 \left( D_{1,q}^{h_1}(z_1) D_{1,\overline{q}}^{h_2}(z_2) +  D_{1,q}^{h_2}(z_2) D_{1,\overline{q}}^{h_1}(z_1)\right),&&
\label{eq:eehhx}
\end{eqnarray}
where it is assumed that both hadrons emerge from different quarks and the scale dependence has been dropped for brevity. This assumption is strictly valid only at leading order \cite{deFlorian:2003cg} and for hadrons that are nearly back-to-back. In order to study its validity, events are analyzed here in three different topologies. When both hadrons are in the same hemisphere as defined by the thrust axis defined below, it is more likely that they emerge from the same parton so that a di-hadron fragmentation function should describe the process. If both hadrons are in opposite hemispheres, the assumption of single-hadron fragmentation for each hadron is more likely. In a third sample, all hadron pairs irrespective of topology are considered. The thrust axis $\hat{\mathbf{n}}$ maximizes the thrust $T$ \cite{thrust}:
\begin{equation}
T \stackrel{\mathrm{max}}{=} \frac{ \sum_h|\mathbf{P^{\mathrm{CMS}}}_h\cdot\mathbf{\hat{n}}|}{ \sum_h|\mathbf{P^{\mathrm{CMS}}}_h|}\quad.
\end{equation}
The sum extends over all detected particles, and $\mathbf{P}^{\mathrm{CMS}}_h$ denotes the momentum of particle $h$ in the CMS.

If the assumption of single-hadron fragmentation holds as described in Eq.~\eqref{eq:eehhx}, the cross sections are then sensitive to favored and disfavored fragmentation depending on the charges and hadron types of the two detected hadrons. For pairs of oppositely charged pions, either both of the hadrons are produced by favored fragmentation or both are produced by disfavored fragmentation; for same-sign pion pairs, one is produced from favored and one from disfavored fragmentation. Consequently, the cross section for same-sign pion pairs is smaller than that for opposite-sign pion pairs if disfavored fragmentation functions are smaller, especially at high $z$ as found in the global fits and expected in models. The reason for the different $z$ dependence originates in the assumption that more quark-antiquark pairs need to be created to arrive at a disfavored hadron, which reduces its large-$z$ possibility. \par
When neglecting strange and charm fragmentation and assuming $SU(2)_F$ isospin symmetry, the ratio of same- over opposite-sign pion pair cross sections reduces, for diagonal $z_1=z_2$ elements, to a simple expression of disfavored and favored light-quark fragmentation functions. Strange and charm fragmentation dilute this simple relation for pions but in a global analysis all yields and all flavors can be treated appropriately. 
This general idea has been formulated in the context of the Collins fragmentation function measurements in $e^+e^-$ and applied there \cite{boer,schweitzer,bellecollins,babarcollins} but has already been considered much earlier in Ref.~\cite{deFlorian:2003cg}.

In the case that two hadrons are detected in the same hemisphere, their production is more likely to arise from the same parton so that di-hadron fragmentation functions (DiFF) should describe their yields theoretically. The formalism for DiFFs was developed initially in Ref.~\cite{diff} and including DGLAP evolution \cite{diffevo1,diffevo2} as summarized in Ref.~\cite{Bacchetta:2008wb}. Their polarized counterparts, sometimes denoted as interference fragmentation functions, have been widely used in SIDIS experiments \cite{Airapetian:2008sk,Adolph:2012nw} and Belle \cite{iff} to access together the quark transverse-spin distribution in the nucleon \cite{Courtoy:2012ry}. In this paper, the individual $z$ dependence of the  unpolarized baseline DiFFs is extracted. \par
It should be noted that the leading-order mapping of single versus di-hadron fragmentation to the opposite versus same hemisphere assignments fails at next-to-leading order, where both types of fragmentation must be considered simultaneously \cite{deFlorian:2003cg}. Due to energy conservation, the momentum of the same-hemisphere di-hadrons should not exceed the total initial momentum of the parton if originating from one parton only. \par
The inclusive cross sections for charged di-hadrons in various topologies as a function of their fractional energies $z_1$ and $z_2$ are extracted in this paper. To evaluate the role of favored and disfavored fragmentation, the ratios between these cross sections for various charge and hadron type combinations are calculated as well. The contributions for different topologies are compared to better understand single versus di-hadron fragmentation. Finally, the cross sections are compared to various Monte Carlo (MC) simulation tunes optimized for different collision systems and energies. \par
Since the corrections are rather similar, a modified version of the di-hadron analysis code is used to extract single-hadron results as a comparison and cross check to the previously published single-hadron cross sections \cite{martin}. As a new result, the previously unpublished single-proton cross sections as a function of $z$ are presented here and compared to the aforementioned MC tunes.
\par
This paper is organized as follows: after a short description of the detector in section I, the raw di-hadron measurement is described in section II before detailing the various corrections necessary to arrive at the final cross sections, their ratios as well as topology dependence. The single-hadron analysis, including the new single-proton results, are presented and compared to MC tunes in section III. We conclude with a summary in section IV.

\section{Belle detector and data selection}

This di-hadron and single-proton cross section measurements are based on data samples of $655\,{\rm fb}^{-1}$ and $159\,{\rm fb}^{-1}$, respectively, 
collected with the Belle detector at the KEKB asymmetric-energy
$e^+e^-$ (3.5~GeV on 8~GeV) collider~\cite{KEKB}
operating at the $\Upsilon(4S)$ resonance (denoted as on-resonance) as well as 60 MeV below for comparison (denoted as continuum).

The Belle detector is a large-solid-angle magnetic
spectrometer that consists of a silicon vertex detector (SVD),
a 50-layer central drift chamber (CDC), an array of
aerogel threshold Cherenkov counters (ACC),  
a barrel-like arrangement of time-of-flight
scintillation counters (TOF), and an electromagnetic calorimeter
comprised of CsI(Tl) crystals (ECL) located inside 
a superconducting solenoid coil that provides a 1.5~T
magnetic field.  An iron flux-return located outside of
the coil is instrumented to detect $K_L^0$ mesons and to identify
muons (KLM).  The detector
is described in detail elsewhere~\cite{Belle}.
Two inner detector configurations were used. A 2.0 cm beampipe with 1 mm thickness 
and a 3-layer SVD were used for the first sample
of $97\,{\rm fb}^{-1}$, while a 1.5 cm beampipe, a 4-layer
SVD and a small-cell inner drift chamber were used to record  
the remaining $558\,{\rm fb}^{-1}$($159\,{\rm fb}^{-1}$ for the single-hadron analysis) \cite{svd2}.  

The primary light- and charm-quark simulations used in this analysis were generated with {\sc Pythia}6.2 \cite{pythia}, embedded into the EvtGen \cite{evtgen} framework, followed by a {\sc Geant}3 \cite{geant} simulation of the detector response. The various MC samples were produced separately for light ($uds$) and charm quarks. In addition, we generated charged and neutral $B$ meson pairs from $\Upsilon(4S)$ decays in EvtGen, $\tau$ pair events with the KKMC \cite{taumc} generator and the {\sc Tauola} \cite{tauola} decay package, and other events with either {\sc Pythia} or dedicated generators \cite{aafh}. 
\subsection{Event and track selection}
Events with at least three reconstructed charged tracks must have a visible energy of charged tracks and neutral clusters above 7 GeV (to remove $\tau$ pair events) and either a heavy jet mass (the greater of the two invariant masses of all particles in a hemisphere) above 1.8 GeV/c$^2$ or a ratio of the heavy jet mass to visible energy above 0.25. 

Tracks must be within 4 cm (2cm) of the event vertex along (perpendicular to) the positron beam axis. Each must have at least three SVD hits and fall within the barrel and full particle-identification (PID) polar-angle acceptance of $-0.511  < \cos\theta_{\mathrm{lab}} < 0.842 $.
The fractional energy of each track must exceed 0.1. (Note that, in this paper, we study fragmentation functions for $z$ above 0.2). This initial fractional energy selection always takes the nominal hadron mass as given by the PID information into account. The requirement of $z > 0.1$ therefore safely accommodates pion-kaon misidentification, which is unfolded in the course of this analysis.

\begin{figure}[ht]
\begin{center}
\includegraphics[width=8cm]{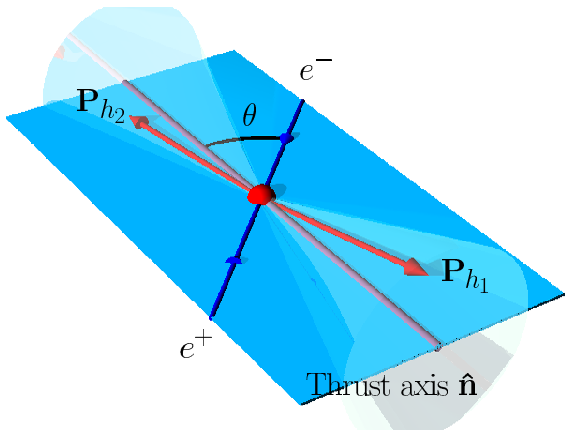} 
\caption{\label{fig:(h)(h)x} Illustration of di-hadron fragmentation where the final-state hadrons are depicted as red arrows, the incoming leptons as blue arrows, and the event plane -- spanned by leptons and thrust axis -- is depicted as a light blue plane. In this case, both hadrons are found in opposite hemispheres defined by the thrust axis, and generally out of the plane, as indicated by the cones.}
\end{center}
\end{figure}

\begin{figure}[ht]
\begin{center}
\includegraphics[width=8cm]{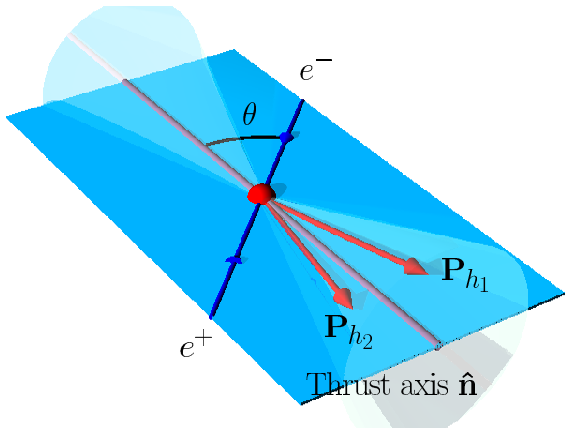} 
\caption{\label{fig:(hh)x} Illustration of di-hadron fragmentation where the final-state hadrons are depicted as red arrows, the incoming leptons as blue arrows, and the event plane -- spanned by leptons and initial quarks/thrust axis -- is depicted as a light blue plane. In this case, both hadrons are found in the same hemisphere as defined by the thrust axis, and generally out of the plane, as indicated by the cones.}
\end{center}
\end{figure}

In addition, in order to study whether two hadrons have likely emerged from the same parton or different partons, the analysis is performed on several different sets by requiring that both hadrons be in opposite hemispheres, the same hemisphere, or anywhere as depicted in Figs.~\ref{fig:(h)(h)x} and \ref{fig:(hh)x}.
For the data sets where a hemisphere assignment is required, the hemispheres are defined by the plane perpendicular to the thrust axis and the thrust must satisfy $T>0.8$. 

\subsection{PID selection}
To apply the PID correction according to the PID efficiency matrices described in Ref.~\cite{martin}, the same selection criteria must be applied to define a charged track as a pion, kaon, proton, electron or muon. The information is determined from normalized likelihood ratios that are constructed from various detector responses.  
If the muon-hadron likelihood ratio is above 0.9, the track is identified as a muon. Otherwise, if the electron-hadron likelihood ratio is above 0.85, the track is identified as an electron. 
If neither of these applies, the track is identified as a kaon by a kaon-pion likelihood ratio above 0.6 and a kaon-proton likelihood ratio above 0.2. Pions are identified with the kaon-pion likelihood ratio below 0.6 and a pion-proton ratio above 0.2. Finally, protons are identified with the inverse proton ratios above with kaon-proton and pion-proton ratios below 0.2. While neither muons nor electrons are considered explicitly for the single and di-hadron analysis, they are retained as necessary contributors for the PID correction, wherein a certain fraction enter the pion, kaon and proton samples under study.   

\section{Di-hadron analysis}
In the following sections, the di-hadron yields are extracted and, successively, the various corrections and the corresponding systematic uncertainties are applied to arrive at the di-hadron differential cross sections $d^2\sigma(e^+e^-\rightarrow h_1h_2X)/dz_1dz_2$. 
\subsection{Binning and cross section extraction}
For the di-hadron cross sections, a ($z_1$, $z_2$) binning is used. We forgo a combined $z$ and invariant-mass binning of the hadron pair; the latter, in particular, is relevant in the {\it same}-hemisphere topology as an unpolarized baseline to the previously extracted interference fragmentation functions \cite{iff} and would have allowed the extraction of individual fragmentation functions for $\rho$, $K^*$, $\phi$ and other resonances. 

The $z_1$ and $z_2$ ranges of 0.2 to 1.0 used in this analysis are each partitioned into 16 equidistant bins. All hadron and charge combinations are treated independently and are merged only after all corrections are applied and after confirming their consistency where applicable (\textit{i.e.,} where the same combinations of fragmentation functions appear, such as $\pi^+\pi^+$ and $\pi^-\pi^-$). This leaves 16 different charge and type combinations for pions and kaons initially, of which six contain irreducible information. 

Furthermore, as mentioned in the introduction, three hemisphere combinations are studied: two hadrons in the same hemisphere, two hadrons in opposite hemispheres, and two hadrons irrespective of hemisphere or thrust cut; these are abbreviated hereinafter as {\em same}, {\em opposite}, and {\em any}, respectively. 

\subsection{PID correction}
\begin{figure*}[ht]
\begin{center}
\includegraphics[width=0.8\textwidth]{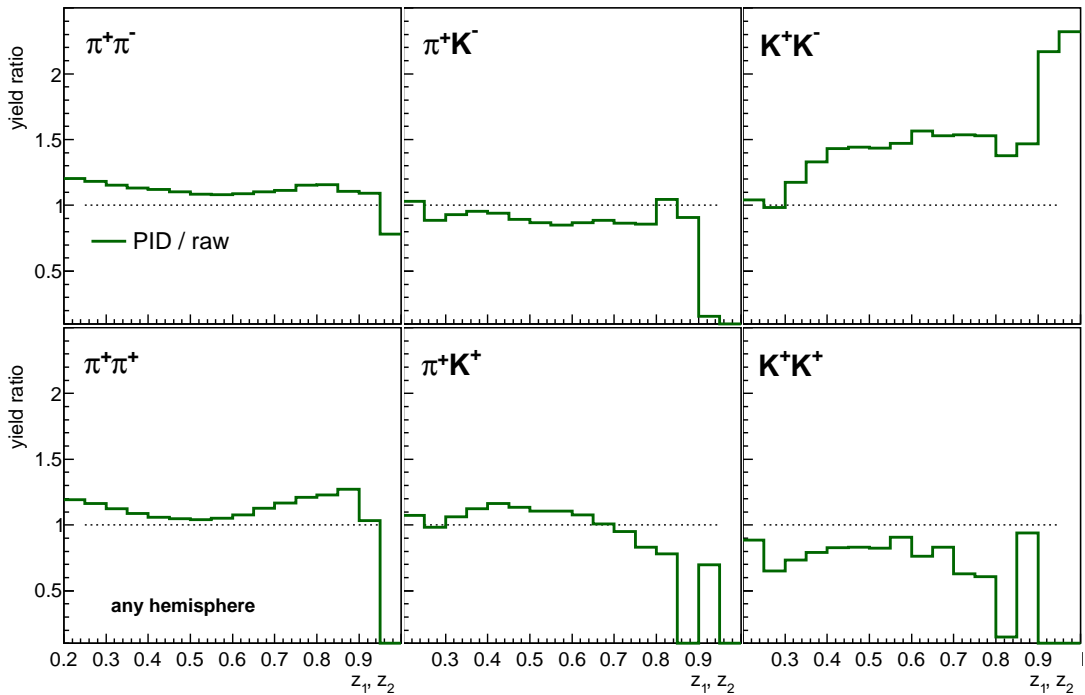} 
\caption{\label{fig:beforeafterpid}(color online) Ratio of yields after and before applying the PID correction for various hadron combinations in {\it any} topology. For brevity, only diagonal ($z_1$ = $z_2$) entries in each two-dimensional matrix are shown. Empty bins are visible where the yields become zero, especially for high-$z$ bins.}
\end{center}
\end{figure*}

As in Ref.~\cite{martin}, the particle misidentification is corrected via inverted $5 \times 5$ particle-misidentification matrices for the five particle hypotheses (pions, kaons, protons, muons, and electrons) for each identified particle, laboratory momentum, and polar angle bin. These matrices are obtained using decays of $D^{*+}$, $\Lambda$ and $J/\psi$ from data where the true particle type is determined by the charge reconstruction and the invariant mass distribution. Occasionally, when too few events are available in the data, the extracted efficiencies are interpolated and/or extrapolated based on the behavior in the generic MC; this occurs particularly at the boundaries of the acceptance.
The matrices are calculated for each of the two-dimensional bins in laboratory momentum and polar angle, with the boundaries of the $17$ bins in momentum at ($0.5, 0.65, 0.8, 1.0, 1.2, ...., 3.0, 3.5, 4.0, 5.0, 8.0$) GeV/$c$ and the boundaries of the 9 bins in $\cos\theta$ at ($-0.511, -0.3, -0.152, 0.017, 0.209, 0.355, 0.435, 0.541,$ $ 0.692, 0.842$). 

In this analysis, the inverted misidentification matrix is applied for each of the identified hadrons by multiplying the respective weights for each hadron being a pion or kaon to obtain the total weight for the di-hadron and any of the four pion-kaon combinations. To confirm the consistency of this treatment, the $D^0$ branching ratios for the pion-pion and kaon-kaon decay channels to the pion-kaon decay channel are compared to the PDG \cite{pdg} values and found to be consistent. We confirm that the total yield of particle pairs is unaffected by this treatment.

The corrected yields are distributed among the ($z_1, z_2$) bins according to the corresponding hadron masses: one identified hadron pair appears in several $z$ bins with the above-determined weights, depending on the particular hadron combination. 
The ratios relative to the uncorrected hadron assignment are displayed in Fig.~\ref{fig:beforeafterpid}, where one can see that the overall corrections are of the order of 20\% to 50\%.

\subsubsection{Uncertainties from the PID correction}
\begin{figure*}[ht]
\begin{center}
\includegraphics[width=0.8\textwidth]{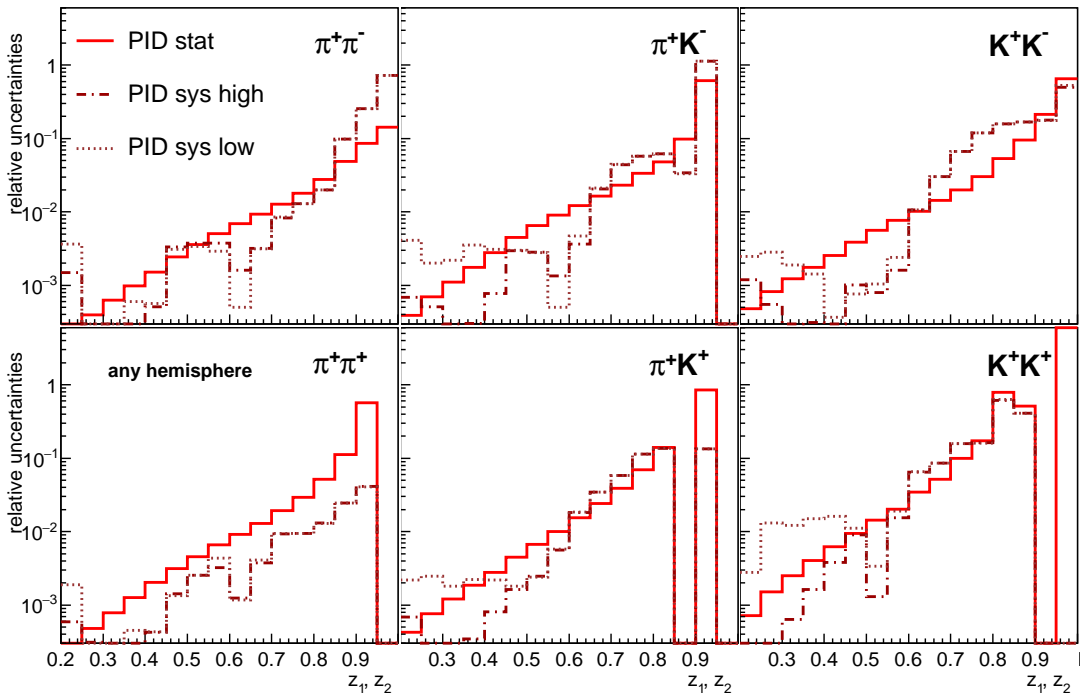}     
\caption{\label{fig:pidsyst}(color online) Statistical (full lines), upper (dashed lines) and lower (red, dotted lines) systematic uncertainties relative to the PID corrected di-hadron yields in {\it any} topology as a function of ($z_1=z_2$); the diagonal bins of each two-dimensional matrix is shown for brevity. Empty bins are visible where the yields become zero, especially for high $z$ bins.} 
\end{center}
\end{figure*}

The uncertainties on the PID matrices are taken into account as uncertainties in the di-hadron yields and propagated through the subsequent corrections. At present, the uncertainties are only assigned individually for each hadron combination, neglecting the correlations between different hadron combinations.
They follow the uncertainties assigned in Ref.~\cite{martin} but take into account the additional complication of having two rather than one hadrons to unfold. To obtain the final uncertainties, the asymmetric uncertainties on the inverted PID matrices are sampled $N$ times with a random generator with Gaussian distributions around the central value (separately above and below this value). From the resulting yields, the 68th percentiles from $N$ samples relative to the central values for each ($z_1,z_2$) bin are taken as the systematic uncertainties on the PID-corrected di-hadron yields due to the PID matrix evaluation and inversion uncertainties. The size of the statistical and systematic uncertainties relative to the PID-corrected di-hadron yields is displayed in Fig.~\ref{fig:pidsyst} for selected hadron combinations. As expected from the overall size of the yields, the statistical precision is best for opposite-sign pion pairs, followed by pion-kaon and then kaon pairs, with the same-sign precision being generally lower. The PID systematic uncertainties are mostly smaller than the statistical uncertainties in almost any bin, with the exception of the lowest $z$ bins.

\subsection{Momentum smearing correction}
The reconstructed fractional momentum $z$ of each hadron may have been smeared from its actual value and therefore must be corrected. For this purpose, the generic MC simulations are used to create two-dimensional histograms with $16\times 16$ bins of generated and reconstructed ($z_1,z_2$) combinations for each of the two hadrons. Only events that are generated and reconstructed within this range of $z$ are considered. Events outside this range are treated in the manner described later in the acceptance correction section. 
The two-dimensional response histograms are created for each hadron and charge combination and for all topology assignments. 
As the PID correction was already applied to the data before the smearing correction, the true particle-type information in the MC is selected for both generated and reconstructed ($z_1,z_2$).
In the smearing matrices, the diagonal elements are dominant for all hadron combinations, as can be expected in such a coarse binning. In the case of the pion-kaon and kaon-kaon combinations, the non-diagonal elements are slightly larger than for the pion-pion case, which indicates that the kaon smearing is slightly larger than that of pions.   
The small off-diagonal components facilitate the inversion of the smearing matrix significantly and a simple, analytically inverted matrix should be sufficient to unfold the di-hadron yields. However, the singular value decomposition (SVD) unfolding method \cite{hoecker} is used as our default to properly unfold the statistical uncertainties and assign systematic uncertainties due to the limited MC statistics, especially for bins far from the diagonal. This also takes into account the possible effects of the different $z$ distributions in data and MC, which will be discussed later. \par
In the SVD unfolding, a regularization parameter $k$ accounts for the lack of statistics in the smearing matrix entries and the shape of the MC spectra. If $k$ is not selected properly, the unfolded yields can be either too biased by the MC spectrum (too small a value for $k$) or can exhibit large fluctuations (too large a value for $k$). Following the procedure of Ref.~\cite{hoecker}, the best regularization parameter is chosen when the index of the regularized vector becomes smaller than unity. Initially, the fluctuations observed in the regularization parameter distribution are large, so that the false choice of too-small $k$ values leads to spurious discrepancies at very high $z$ between distributions of the same physics content that were consistent before unfolding. After smoothing the $k$ distributions, the expected behavior is markedly better (\textit{i.e.,} a relevant exponentially falling contribution and an irrelevant flat contribution due to MC statistical fluctuations) and the selection of the regularization parameter is considered more reliable. 

\begin{figure*}[ht]
\begin{center}
\includegraphics[width=0.8\textwidth]{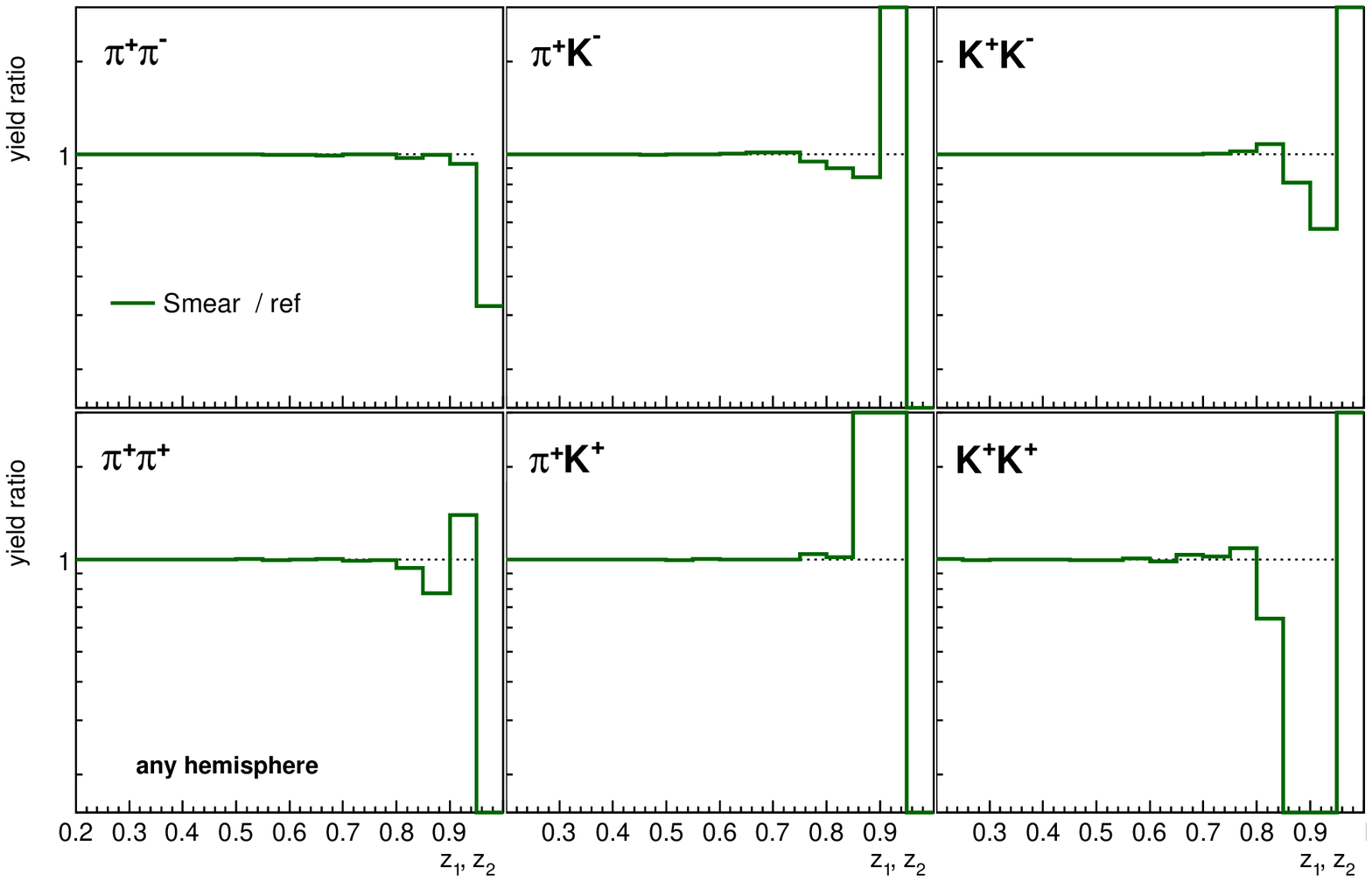} 
\caption{\label{fig:beforeaftersmearing}(color online) Ratio of yields after and before applying the smearing correction for various hadron combinations without hemisphere assignment. For brevity, only diagonal ($z_1$ = $z_2$) entries in each two-dimensional matrix are shown.}
\end{center}
\end{figure*}

The smearing is corrected in all data samples after the PID correction is applied and before the non-$q \bar{q}$ events are removed. The final before/after ratio plots are displayed in Fig.~\ref{fig:beforeaftersmearing} as a function of ($z_1,z_2$). Apart from the highest ($z_1,z_2$) bins, where the corrections get large, the yield ratios are close to unity.

All uncertainties prior to the smearing-unfolding (PID and statistical uncertainties) are unfolded as well, resulting in the respective covariance matrices. The covariance matrix due to the MC statistics itself and the differences with an analytic unfolding (\textit{i.e.,} application of the inverted response matrix) are assigned as systematic uncertainties related to the unfolding.  
\subsection{Non-$q \bar{q}$ background correction}
Various QED processes can produce hadronic final states that contribute to our di-hadron yields and must be removed. Apart from particle misidentification, which has been addressed already, $e^+e^- \rightarrow \mu^+\mu^-,\ e^+e^-e^+e^-,\ e^+e^-\mu^+\mu^-$ and Bhabha scattering processes cannot contribute to hadronic final states, as has been verified in MC simulations. 
The processes that do produce hadron pairs are either QED processes having partons created initially, such as two-photon processes $e^+e^- \rightarrow  e^+e^- u\overline{u},e^+e^-d\overline{d},e^+e^-s\overline{s}$ and $e^+e^-c\overline{c}$, or via decays such as from $e^+e^- \rightarrow  \tau^+\tau^-$. Hadrons from these processes are not produced directly via $e^+e^-\rightarrow q\overline{q}$ and hence should not be included in our extracted di-hadron cross sections. 
Similarly, resonant $\Upsilon(4S)$ production and subsequent decays into neutral or charged $B$ meson pairs create pion and kaon pairs that must be removed (the non-resonant $e^+e^-\rightarrow b\bar{b}$ process does not contribute \cite{Santel:2015qga}). 
The direct production of quark-antiquark pairs in electron-positron annihilation $e^+e^-\rightarrow u\bar{u},\ d\bar{d},\ s\bar{s}$ and $c\bar{c}$ is treated as signal in this section, while weak decays in these continuum processes will be treated later.\par  

\begin{figure*}[t]
\begin{center}
\includegraphics[width=0.85\textwidth]{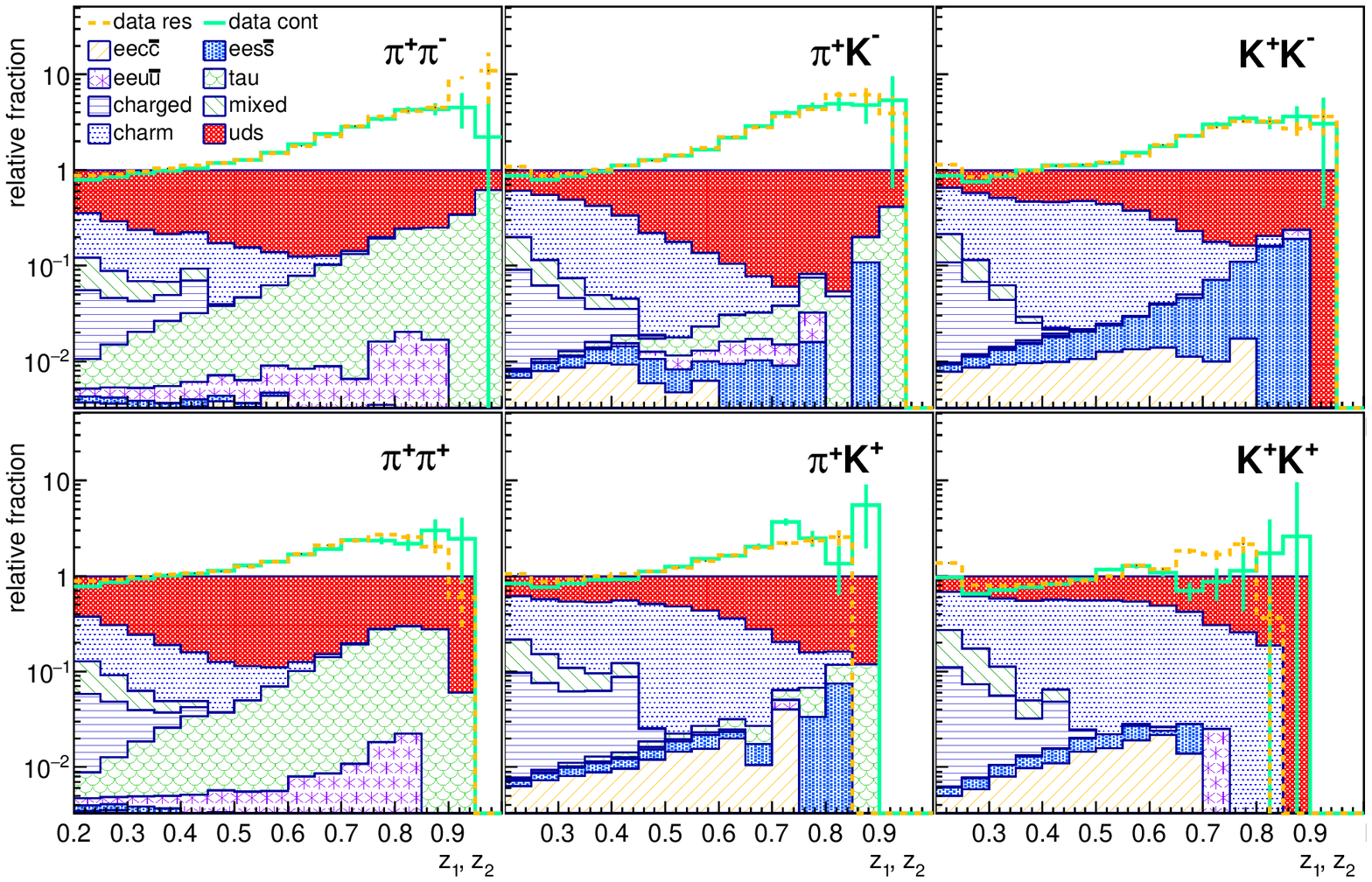}
\caption{\label{fig:zqedcontpid0_mix4}(color online) Fraction of hadron pairs in {\it any} topology as a function of ($z_1,z_2$) originating from various sub-processes. The individual relative contributions are displayed from top to bottom for $uds$ (red filled area), charm (blue, dotted area), mixed ($\Upsilon (4S)\rightarrow B^0\overline{B}^0$, dark-green, hatched area) and charged ($\Upsilon (4S)\rightarrow B^+B^-$, violet, horizontally hatched area), $\tau$ pair (light green, scaled area), $eeu\bar{u}$ (purple, starred area), $ees\bar{s}$ (light blue, dotted area) and $eec\bar{c}$ (orange hatched area) events. 
Also, for comparison, the continuum (green, solid lines) and on-resonance (orange, dotted lines) data are shown. For brevity, only diagonal ($z_1$ = $z_2$) entries in each two-dimensional matrix are shown.}
\end{center}
\end{figure*}

Figure \ref{fig:zqedcontpid0_mix4} shows the relative fractions of all these processes for selected hadron pairs in the {\it any} di-hadron topology. Due to the large branching fraction of the single-prong $\tau$ decay for at least one of the $\tau$ leptons, $\tau$ processes are the dominant background for pions from small to especially large fractional energies where the single hadron inherits a large fraction of the $\tau$ momentum. The other non-$q \bar{q}$ processes generally play a minor role with contributions less than a few percent, with the exception at high $z$ where the two-photon process $ee (u\bar{u},d\bar{d})$ contributes several percent to pion pairs. 
Resonant $\Upsilon(4S)$ production either in charged or neutral ($CP$-mixed) $B$ meson decays contributes a few to about 10\% and it vanishes when one fractional energy approaches 0.5 due to the additional decays needed to produce pions and kaons. 
The distributions for other di-pion topologies are similar except that the additional thrust requirement removes nearly all $\Upsilon(4S)$ decays. 
For same-sign di-pions, the $\tau$ contribution is substantially smaller as the single-prong decays of oppositely charged $\tau$ create predominantly oppositely-charged pions. For {\it same}-hemisphere di-pions, the single-prong $\tau$ decays cannot contribute and consequently the relative $\tau$ contributions are below 10-20\% everywhere.  

For kaon-related di-hadron combinations, the overall non-$q \bar{q}$ contributions are as small as for di-pions, but $ee s \overline{s}$ and $ee c \overline{c}$ are more important. In addition, the $\tau$ decays do not play a substantial role due to the suppressed kaonic decays. Charm decays generally produce more CKM-favored \cite{KM} kaons than CKM-suppressed pions. This results in a generally larger fraction of charm events contributing to the pion-kaon and kaon-kaon cross sections: up to 60\% for kaon pairs at the lowest $z$, with a similar fall-off as for pion pairs.  Similarly, $\Upsilon (4S)$ decays favor kaons over pions and thus their fractions are as high as 20\% (summed), rapidly disappearing at higher $z$.     

Assuming that the non-$q \bar{q}$ and $\Upsilon$ MC reliably describe the data, the background contributions are directly subtracted from the data distributions. In this way, we avoid introducing further uncertainties due to the shape of the $udsc$ MC. As all these processes are QED and $\Upsilon(4S)$ processes, they are very well understood at the theory level. The yield uncertainty is 1.4\% for the $e^+e^-\rightarrow \tau^+\tau^-$ process \cite{taumc} but is substantially larger for the two-photon processes due to associated production, which is not taken into account in the current two-photon simulations. A factor of four relative to the nominal yield has been assumed for the latter \cite{uehara}. For the systematic uncertainties due to the non-$q \bar{q}$ background correction, these overall uncertainties as well as the statistical uncertainties in the non-$q \bar{q}$ MC, are taken into account.  

\subsection{Preselection and acceptance correction}
The preselection and acceptance correction is divided into three separate terms, motivated by the different sources of corrections and to better expose their individual effects. The first takes into account the effect on the reconstruction within the specified acceptance selection, mostly due to the preselection criteria and decays in flight; the second treats the losses outside the barrel acceptance; and the third takes into account potential losses as $|\cos\theta|$ approaches unity, which are not properly described in the generic MC. 
\subsubsection{Reconstruction efficiency within the barrel acceptance}
The first correction incorporates generated hadron pairs within the barrel geometry that do not get reconstructed. As particle identification, non-$q \bar{q}$ removal and smearing have already been applied, the reconstructed events are considered based on the generic reconstructed $udsc$ MC information, but taking the MC-truth particle type and momenta instead of the reconstructed values. The correction factor is calculated as the ratio of reconstructed to generated events per ($z_1,z_2$) bin for each hemisphere assignments, hadron types, and charge combinations. This correction takes also into account the events that were initially smeared out of or into the $z$ range considered for this analysis. 

The efficiencies are relatively flat at around 70\% and only drop substantially at higher $z$. This is similar to the behavior noticed in Ref.~\cite{martin}, where it was found to be mostly due to the preselection criteria, especially the heavy jet mass restriction that disfavors high-$z$ events where the hadrons naturally have to be more aligned with the thrust axis as little other energy remains. Also, the minimum track requirement of three disfavors very high $z$ hadron pairs, where for the same reason the multiplicity is small. 

\subsubsection{Acceptance outside the barrel region}
A certain fraction of di-hadrons are not reconstructed because at least one of the hadrons is outside of the barrel acceptance. This fraction is evaluated by comparing the generated MC within the barrel acceptance (\textit{i.e.,} including the acceptance selection criteria) with the generated MC without the acceptance requirement. This acceptance fraction is around 70\% and rather flat as a function of ($z_1,z_2$), increasing slightly at very high fractional energies. 

The only systematic uncertainties related to these acceptance corrections (both within and outside the barrel region) originate from the statistical uncertainty of the fractions within the acceptance. These uncertainties are rather moderate in comparison to all other systematic and statistical uncertainties that are aggregated in section \ref{sec:syst}. 

\subsubsection{Large $|\cos\theta|$ region}
The generated MC does not necessarily reproduce the hadron distributions well for very forward or backward polar angles in the CMS. Ideally, the hadron polar angular distributions should resemble those of the initially produced quark-antiquark pairs and thus follow a $(1+\cos^2\theta)$ dependence, neglecting the small linear dependence due to $\gamma-Z$ interference. The lower the fractional energy, the less pronounced this behavior: this is due to the additional smearing by the transverse momentum generated in the fragmentation process. While such a behavior is roughly visible at smaller polar angles, the distributions rapidly drop off at higher polar angles as if some remaining acceptance cut is still present. 
As a consequence, the previous acceptance and efficiency corrections are not complete and need to be further corrected for this effect. 
As the dependence at smaller polar angles is well described by the expected parabola, this function is used to fit the MC and compare the areas below the fit-result curve and the actual histograms. In principle, this treatment should be independent for the two hadrons and can be applied by multiplication of the two individual correction factors.
An expected increase of the correction with increasing $z_2$, due to higher-$z$ tracks being more collimated and thus closer to the partonic polar angular dependence, has been confirmed. The overall effect of this last acceptance correction is on the order of a few percent.

\begin{figure*}[ht]
\begin{center}
\includegraphics[width=0.85\textwidth]{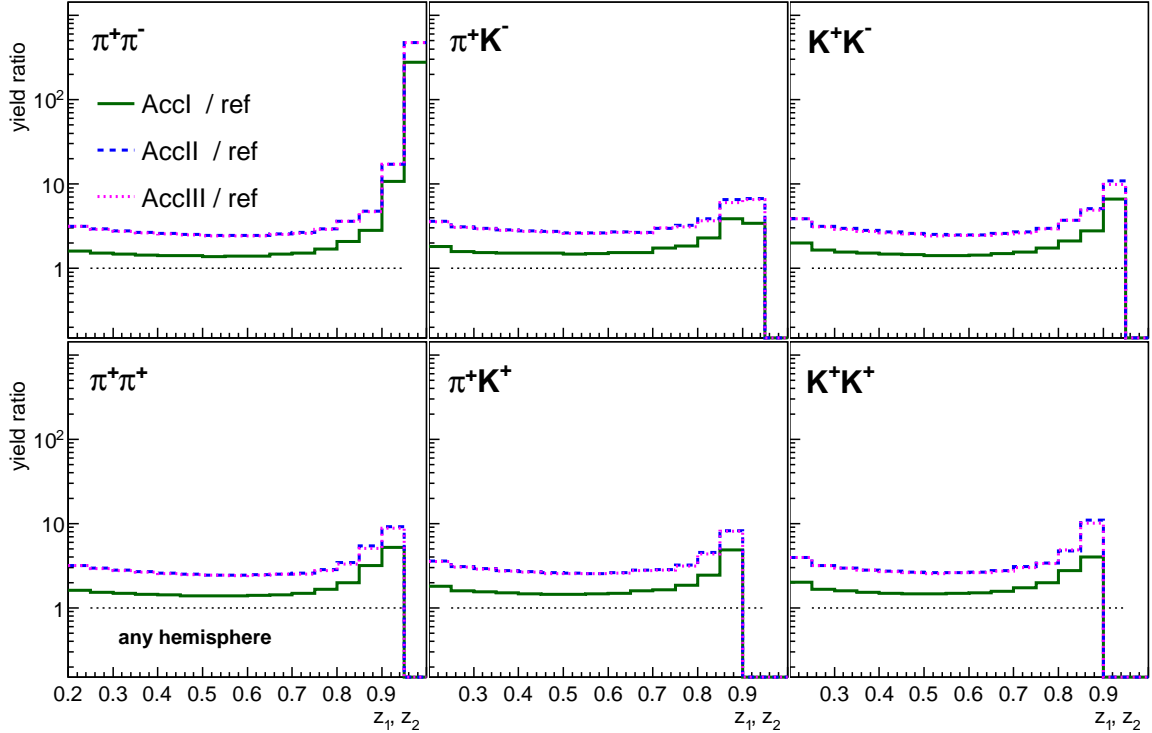} 
\caption{\label{fig:acceptance}(color online) Yield ratios after successively applying all acceptance and efficiency corrections (labeled AccI to AccIII as the three acceptance corrections discussed in the text) relative to the reference yields before (non-$q\bar{q}$ removal) for various hadron combinations without hemisphere assignment. For brevity, only diagonal ($z_1$= $z_2$) entries in the two-dimensional matrices are shown.}
\end{center}
\end{figure*}

The effect of all three acceptance and efficiency corrections is summarized in Fig.~\ref{fig:acceptance}, where the ratios of the di-hadron yields before and after the corrections are displayed. The overall effect amounts to between two times the initial yields at moderate ($z_1,z_2$) and more than ten times at very high $z$ (where the event preselection correction dominates).

\subsection{Weak decays}
Generally, fragmentation functions are only defined for hadrons produced by QCD processes and decays and so any weak decays should be removed. In practice this is only possible ---if at all--- with the help of MC and not entirely reliable. Therefore, many fragmentation results do not exclude weak decays or only those experimentally detectable such as those of $\Lambda$ baryons and neutral kaons.  
The approach taken here is to provide results that either contain all weak decays or completely remove them with the help of MC.  
Every $c\bar{c}$ event undergoes at least one weak decay to produce a pion or kaon. However, in the fragmentation process, various quark-antiquark pairs are created and consequently pions and kaons can be created that did not originate directly from the decays of charmed hadrons. The only way to separate them is by following the parents of each final state hadron in the MC to either a gluonic string, which corresponds to the absence of a weak decay, or a hadron with a different, non-light valence flavor. In the latter case, a weak decay was present and this hadron would have to be removed. 
The difficulty is rapidly (algorithmically) determining this information for a given hadron type. In the di-hadron analysis, it can be argued that the chance of at least one of the two hadrons being from a weak decay is much higher for charm events and that removing all charm events is a valid approximation. However, this needs to be tested.

The MC history of each hadron is studied to find weak decays. The heaviest flavor of each particle in the decay chain is selected and compared to the mother particles. If the decay chain ends at a string without a change in its heaviest flavor, no weak decay is present. If the flavor does change, a few strong decays need to be vetoed before asserting the presence of a weak decay. Examples are various vector mesons and other excited states with nonzero strangeness where the strangeness is retained in a lower mass state, such as $K^*\rightarrow K \pi$. Also, various $s\bar{s}$ and $c\bar{c}$ resonances need to be excluded as they also decay strongly despite the Zweig rule \cite{zweig}. 

\begin{figure*}[t]
\begin{center}
\includegraphics[width=0.85\textwidth]{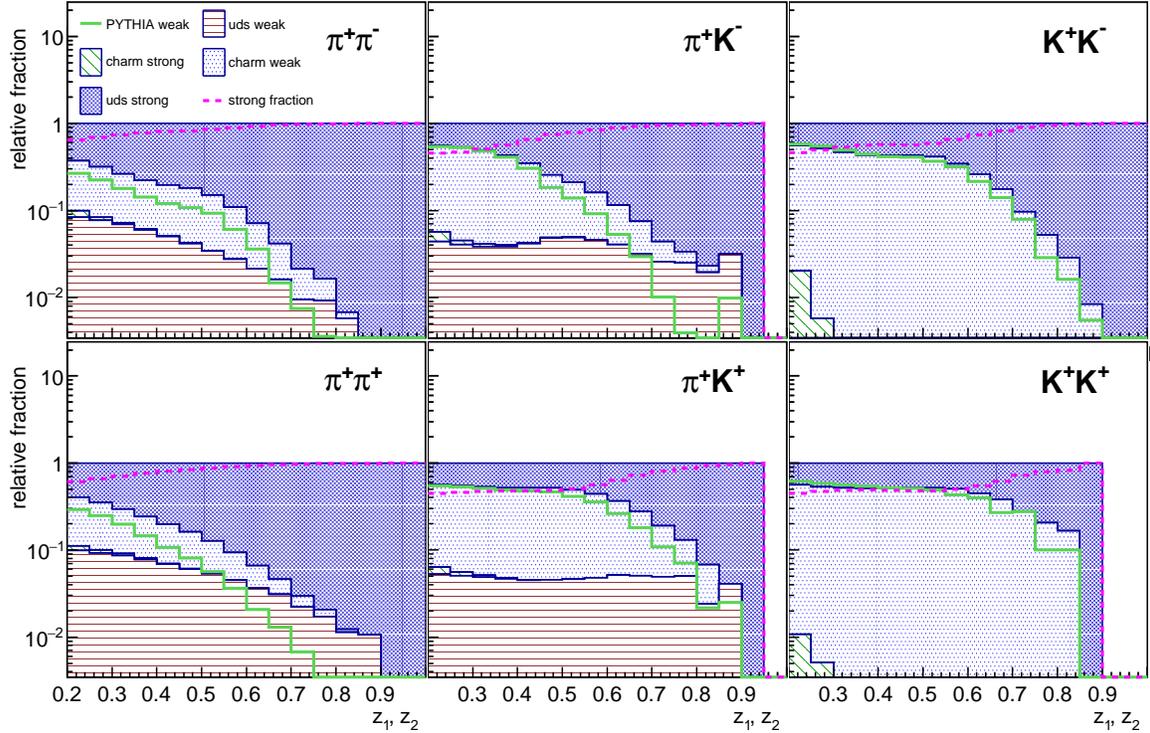}

\caption{\label{fig:zweakratiop_mix4}(color online) Fraction of hadron pairs in the {\it any} topology as a function of ($z_1,z_2$) originating from weak and strong decays. The individual relative contributions are displayed from top to bottom for strong $uds$ decays (purple, dark filled area), weak charm decays (blue dotted area), strong charm decays (dark-green, negative hatched area) and weak $uds$ decays (red, horizontal striped area). The strong decay fractions are also displayed as dashed magenta lines. Also the weak decay fractions for $udsc$ MC using the default {\sc Pythia} settings are indicated by the dark-green, solid lines. For brevity, only diagonal ($z_1$ = $z_2$) entries in each of the two-dimensional matrices are shown.}
\end{center}
\end{figure*}

The overall weak- and strong-decay fractions as a function of ($z_1,z_2$) are shown in Fig.~\ref{fig:zweakratiop_mix4} for the main particle combinations within the {\it any} topology. Similar results are obtained for the other two topologies. It should be noted that the assumption of charm events creating only weak-decay pions and kaons is almost fulfilled in the procedure mentioned above but that a small fraction of charm di-hadron events nevertheless originates in strong decays. Overall, the fraction of strong decays dominates in all ($z_1,z_2$) bins for pion pairs, while the higher fraction of charm events results in larger weak fractions for pion-kaon and is even more pronounced in kaon-kaon combinations. In all cases, the weak fractions drop with $z$ as the additional decays soften the spectrum.  

These strong/weak fractions are model-dependent statements as the fragmentation process is only approximated in {\sc Pythia}. 
Furthermore, the absolute size of weak decays within {\sc Pythia} depends also on the fragmentation settings. The uncertainties due to these effects are evaluated by comparing the generic Belle MC to the {\sc Pythia} default settings in the MC. The strong fractions for the {\sc Pythia} default MC are given in the plots as well for comparison. As can be seen, they are rather similar but are generally slightly lower at high $z$ where the generally harder spectra in the default {\sc Pythia} settings allow for slightly more weak decays to be present. The differences are assigned as a systematic uncertainty for the cross sections that have the weak processes removed.

\subsection{ISR correction}
\begin{figure*}[t]
\begin{center}
\includegraphics[width=0.8\textwidth]{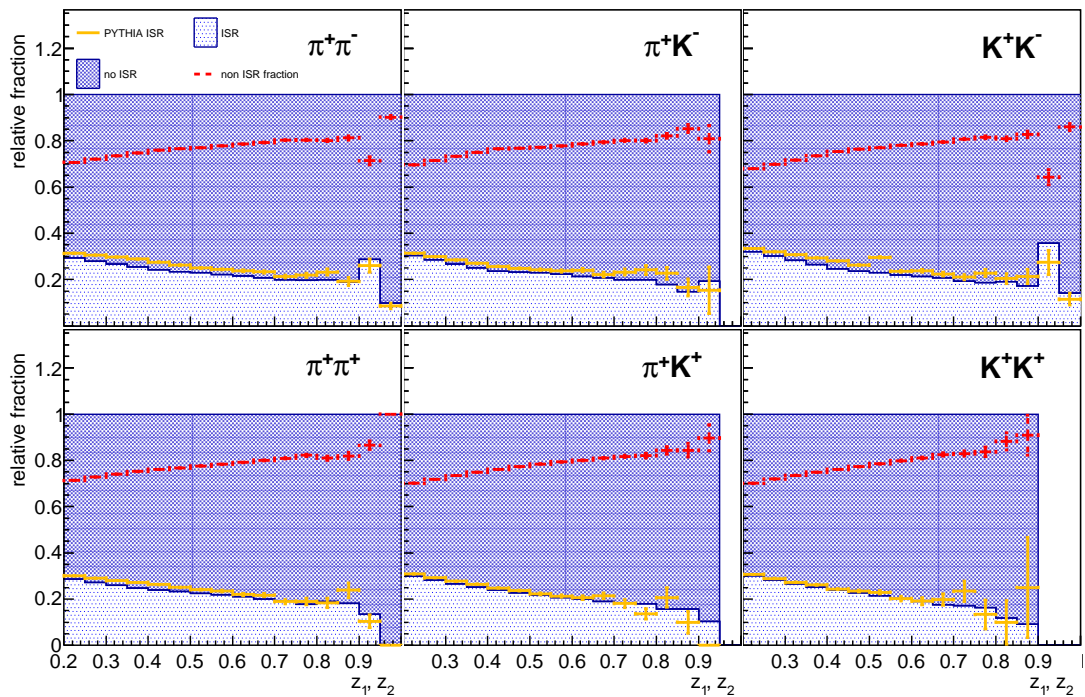}
\caption{\label{fig:isrzfraction_pid0_mix4}(color online) Relative Fractions of hadron pairs in the {\it any} topology as a function of ($z_1,z_2$) originating from ISR or non ISR events. The individual relative contributions are displayed from top to bottom for non ISR events (energy loss less than 0.5\%, filled purple area) and ISR events (blue, dotted area) from generated generic $udsc$ MC. The non-ISR fraction is also shown (red, dashed line). The relative ISR fraction for the default {\sc Pythia} MC is also shown (orange, solid lines). For brevity, only diagonal ($z_1$ = $z_2$) entries in each of the two-dimensional matrices are shown.}
\end{center}
\end{figure*}

Initial-state radiation reduces the CMS energy of the produced quark-antiquark pair. Consequently, the fractional energies calculated relative to the nominal CMS energy are not correct. This can alter the shape of the actual $z$ dependence of the fragmentation functions and also invalidates pQCD calculations evaluated at the nominal CMS energy.
The correction procedure relies on the strategy applied in Ref.~\cite{martin} for the single-hadron cross sections. The events are classified according to their difference from the nominal CMS energy; the events with a CMS energy below 99.5\% of the nominal energy are removed. 
Ideally, one would want to observe the initial-state radiation directly in the reconstructed data; however, most photons are in the very forward and backward regions, outside the Belle acceptance. Instead, generated MC data are used to directly identify ISR photons and remove their energies from the total CMS energy. In the MC, the ISR photons are identified by having their mother particles be an initial-state lepton.
The fraction of such events depends on the fractional energies of the two final-state hadrons. If a large amount of the energy is removed by the photons from the produced quark-antiquark system, very high fractional energies with respect to the nominal $\sqrt{s}$ are inaccessible. Therefore the fraction of non-ISR events (\textit{i.e.,} less than 0.5\% CMS energy loss) increases with increasing fractional energies. This is indeed the case as can be seen in Fig.~\ref{fig:isrzfraction_pid0_mix4} for the {\it any} topology hadron pairs (and similarly for the other two topologies). 
The events are then corrected by this fraction to obtain the ISR-free differential cross sections at the nominal center-of-mass energy. 
Since the ISR fraction depends on the fractional energies of the hadrons, the ($z_1,z_2$) distribution shape of the MC simulation enters in the ISR correction. To address the dependence of the ISR correction on the shape in the MC, an alternative MC is used for comparison and the differences in the extracted di-hadron cross sections is assigned as systematic uncertainties. These fractions are also shown in Fig.~\ref{fig:isrzfraction_pid0_mix4}. The ISR fractions are found to be consistent within the limited precision for both {\sc Pythia} settings.  

\begin{figure*}[ht]
\begin{center}
\includegraphics[width=0.85\textwidth]{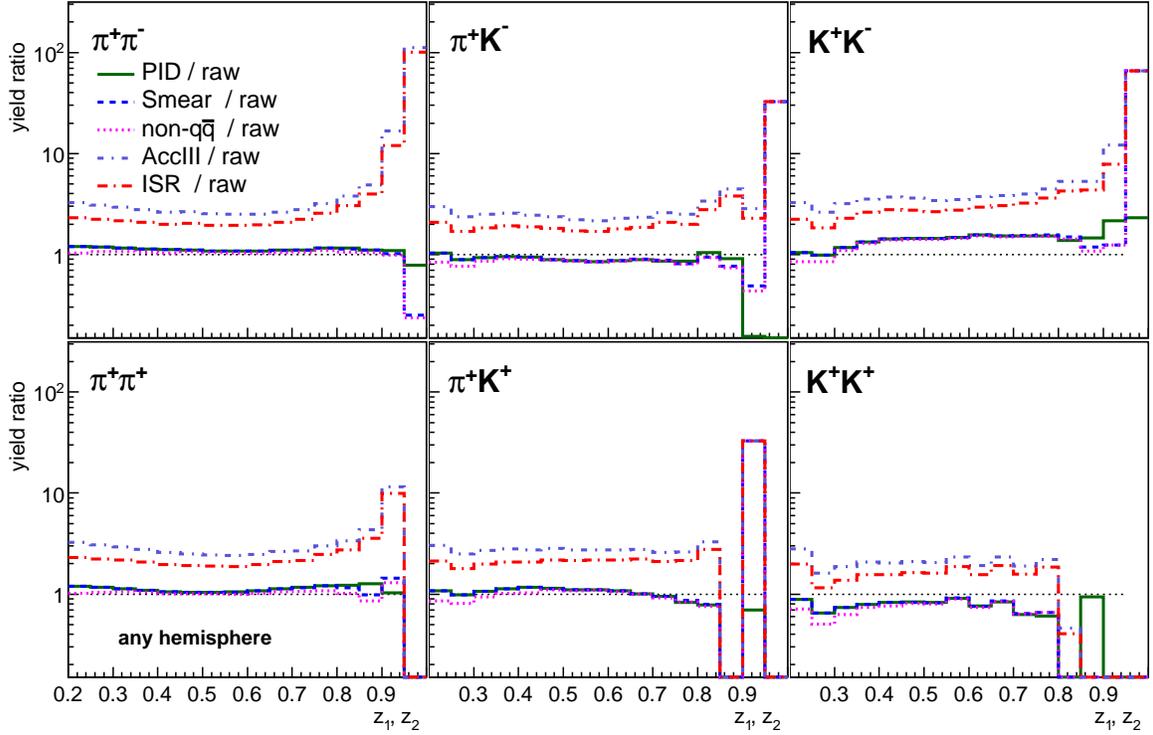} 
\caption{\label{fig:zzbeforeall}(color online) Ratio of yields after various corrections relative to the raw yields for the main hadron combinations without hemisphere assignment. From top to bottom, the ratios after acceptance correction (purple, dash-dotted line), initial state radiation (red, long dash-dotted line), PID correction (dark green, full lines), smearing correction (blue, dashed line) and non-$q\bar{q}$ removal (magenta, dotted lines) are shown. For brevity, only diagonal ($z_1$ = $z_2$) entries in each of the two-dimensional matrices are shown.}
\end{center}
\end{figure*}

The total impact of all corrections from the particle misidentification to the correction for ISR can be seen successively in Fig.~\ref{fig:zzbeforeall} for the main hadron combinations without a hemisphere assignment. The overall correction to the raw yields is substantial, predominantly due to the necessary acceptance corrections. They are comparable for most ($z_1,z_2$) bins but rise at the highest $z$ bins due to the acceptance and smearing corrections. 

\subsection{Consistency checks and total systematic uncertainties\label{sec:syst}}
To confirm the consistency of the results, various tests are performed. For example, the dependence on the data-taking periods is studied; after taking into account variations in acceptance and reconstruction efficiency, the cross sections are consistent within several percent between different periods and no additional systematic uncertainty is assigned. 
In another study, we compare the data recorded at the $\Upsilon(4S)$ resonance with the smaller off-resonance data sample. After removal of the $\Upsilon(4S)$ decay contributions in the non-$q \bar{q}$ correction, the results from both collision energies are consistent.  
In yet another set of comparisons with the same physics-related information, such as charge conjugation of both particles ($\pi^+\pi^+ \leftrightarrow \pi^-\pi^-$, etc.) or (random) hemisphere assignments ($\pi^-K^+ \leftrightarrow K^+\pi^-$), no systematic differences beyond the assigned uncertainties are found.

\begin{table*}[htb]
\begin{center}
\caption{\label{tab:systall} Systematic and statistical uncertainty contributions for the main hadron combinations in the {\it any} topology integrated over the entire ($z_1,z_2$) range. The uncertainties due to the luminosity and track reconstruction are additional global uncertainties. }
\begin{tabular}{c c c c c c c }
 & $\pi^+\pi^-$& $\pi^+\pi^+$ & $\pi^+K^-$& $\pi^+K^+$ & $K^+K^-$& $K^+K^+$\\ \hline
Statistical& $8.71\cdot 10^{-05}$& $1.11\cdot 10^{-04}$& $1.56\cdot 10^{-04}$& $1.73\cdot 10^{-04}$& $1.83\cdot 10^{-04}$& $3.31\cdot 10^{-04}$\\ 
\hline  PID& $9.61\cdot 10^{-04}$& $4.78\cdot 10^{-04}$& $2.09\cdot 10^{-03}$& $1.85\cdot 10^{-03}$& $2.57\cdot 10^{-03}$& $3.06\cdot 10^{-03}$\\ 
 Smearing& $6.31\cdot 10^{-05}$& $3.42\cdot 10^{-05}$& $3.92\cdot 10^{-04}$& $2.07\cdot 10^{-05}$& $6.69\cdot 10^{-05}$& $2.75\cdot 10^{-04}$\\ 
non-$q\bar{q}$ & $6.07\cdot 10^{-04}$& $6.30\cdot 10^{-04}$& $1.03\cdot 10^{-03}$& $9.98\cdot 10^{-04}$& $1.14\cdot 10^{-03}$& $1.88\cdot 10^{-03}$\\ 
 Acceptance & $1.16\cdot 10^{-03}$& $1.32\cdot 10^{-03}$& $2.04\cdot 10^{-03}$& $2.14\cdot 10^{-03}$& $2.24\cdot 10^{-03}$& $3.65\cdot 10^{-03}$\\ 
ISR & $3.66\cdot 10^{-04}$& $4.13\cdot 10^{-04}$& $5.97\cdot 10^{-04}$& $6.09\cdot 10^{-04}$& $7.12\cdot 10^{-04}$& $1.03\cdot 10^{-03}$\\ 
\hline Combined systematics& $1.86\cdot 10^{-03}$& $1.71\cdot 10^{-03}$& $3.82\cdot 10^{-03}$& $4.38\cdot 10^{-03}$& $4.21\cdot 10^{-03}$& $5.28\cdot 10^{-02}$\\ 
\hline Luminosity & \multicolumn{6}{c}{$1.4\cdot10^{-02}$}\\ 
\hline Track reconstruction & \multicolumn{6}{c}{$0.7\cdot 10^{-02}$}\\ 
\end{tabular}
\end{center}
\end{table*}

\begin{figure*}[htb]
\begin{center}
\includegraphics[width=0.85\textwidth]{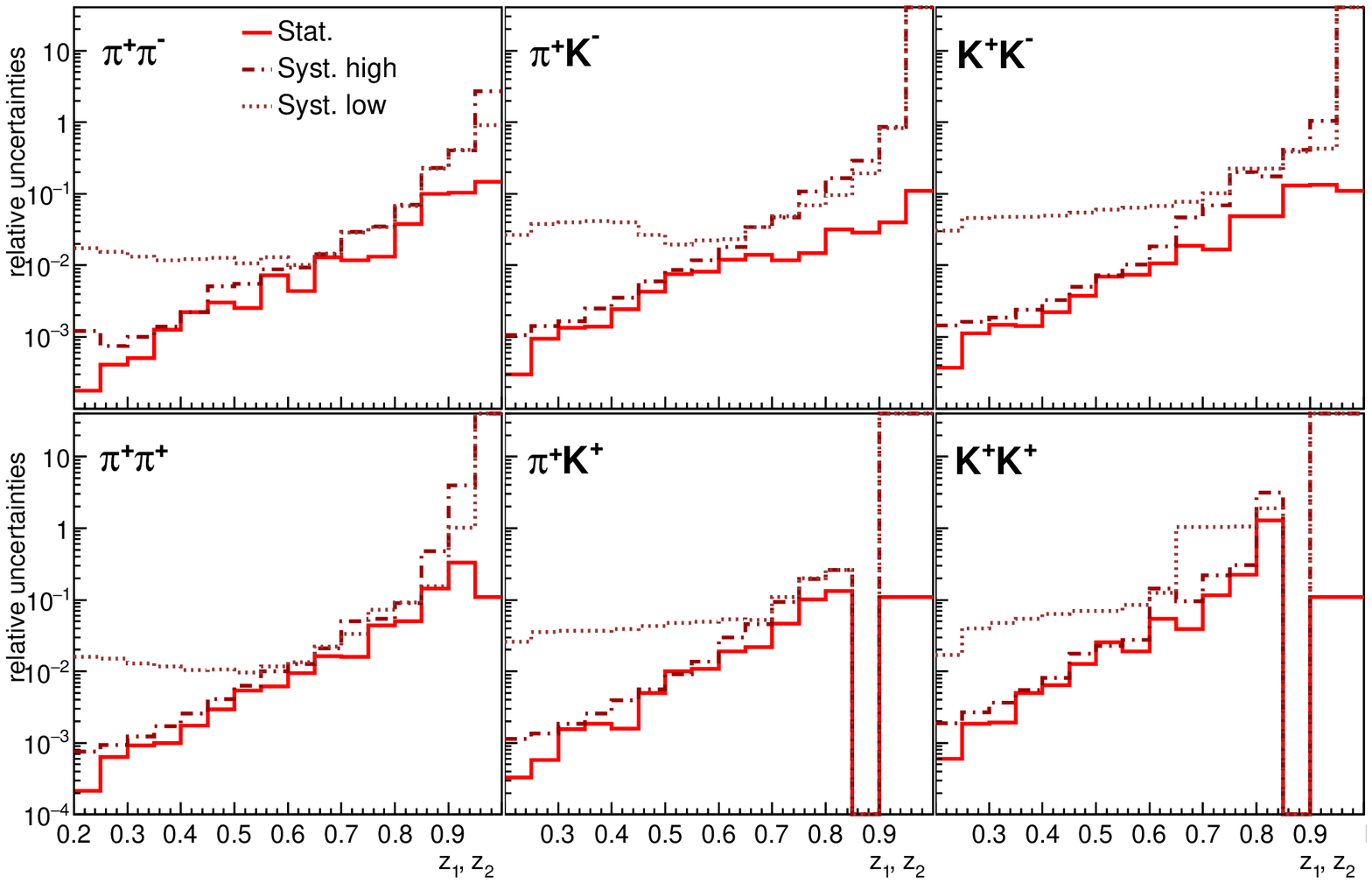}

\caption{\label{fig:systall_mix4_sum}(color online) Relative (asymmetric) systematic (lower uncertainties - dashed lines; upper uncertainties - dash-dotted lines) and statistical uncertainties (full lines) for the most relevant hadron pairs in the {\it any} topology as a function of ($z_1,z_2$). For brevity, only the diagonal bins ($z_1 = z_2$) are shown.}
\end{center}
\end{figure*}

All diagonal systematic uncertainties are summed in quadrature. The total relative systematic uncertainties along with the statistical uncertainties are displayed in Fig.~\ref{fig:systall_mix4_sum} for the relevant hadron pairs without topology assignment for diagonal ($z_1,z_2$) bins and in Table \ref{tab:systall} for the entire measurement range. This measurement is limited almost everywhere by the systematic uncertainties, for which the dominant contributions arise from the smearing correction except at high $z$ where the rapidly falling MC precision contributes comparably. With increased MC data the systematic uncertainties could be reduced to the level of the statistical uncertainties and be dominated by the smearing correction. 
Additionally, there are global scale uncertainties due to the luminosity measurement (1.4\%) and the track reconstruction (2$\times$0.35\%) are not shown.  
  
\subsection{Results}
\begin{figure*}[htb]
\begin{center}
\includegraphics[width=0.9\textwidth]{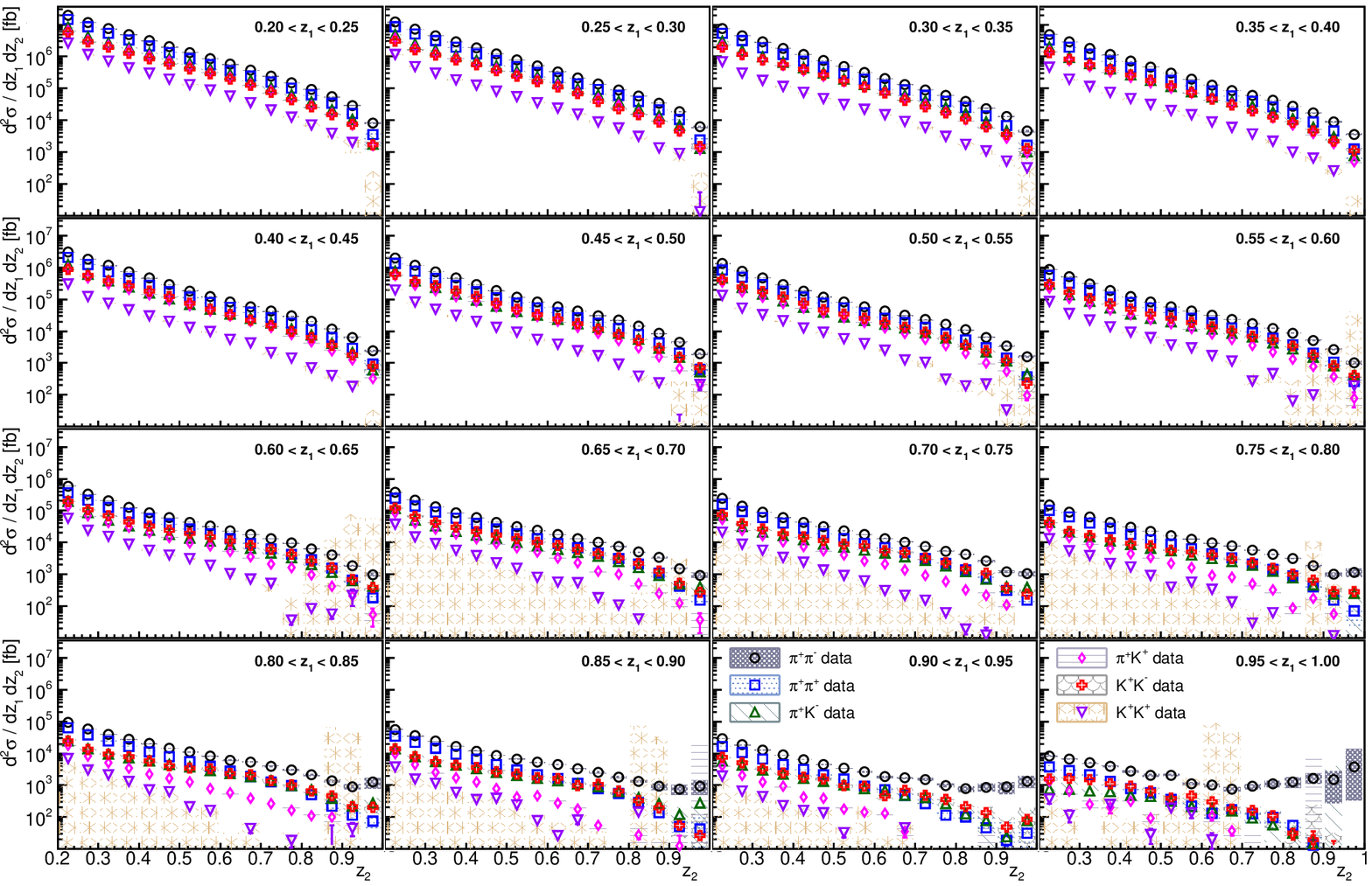}

\caption{\label{fig:yieldall_mix4_pid0_1_4_12}(color online) Differential cross sections for $\pi^+\pi^-$ (black circles), $\pi^+\pi^+$ (blue squares), $\pi^+K^-$ (green triangles), $\pi^+K^+$ (purple diamonds), $K^+K^-$ (red crosses) and $K^+K^+$ (violet downward triangles)  pairs in the {\it any} topology as a function of $z_2$ for the indicated $z_1$ bins. The error boxes represent the systematic uncertainties.}
\end{center}
\end{figure*}

The final cross sections for the main hadron-pair combinations are presented in Fig.~\ref{fig:yieldall_mix4_pid0_1_4_12} for all ($z_1,z_2$) bins and without topology assignment. The results shown here and elsewhere include weak decays unless otherwise noted.
As expected, the opposite-sign pion pairs have the largest cross sections at all $z$ combinations, followed by the same-sign pion pairs. However, the oppositely charged pion-kaon and kaon-kaon combinations seem to be of similar magnitude or even larger than the same-sign pions at higher $z$, which might be explained by the potentially larger favored fragmentation combination from strange-quark pairs. Same-sign kaon pairs have the lowest cross sections in general, with the relative differences from the other combinations increasing at increasing $z$. As at least one kaon in this case needs to be produced from disfavored fragmentation, the additional strangeness suppresses the cross sections beyond that for the disfavored pion fragmentation functions. 

\begin{figure*}[htb]
\begin{center}
\includegraphics[width=0.9\textwidth]{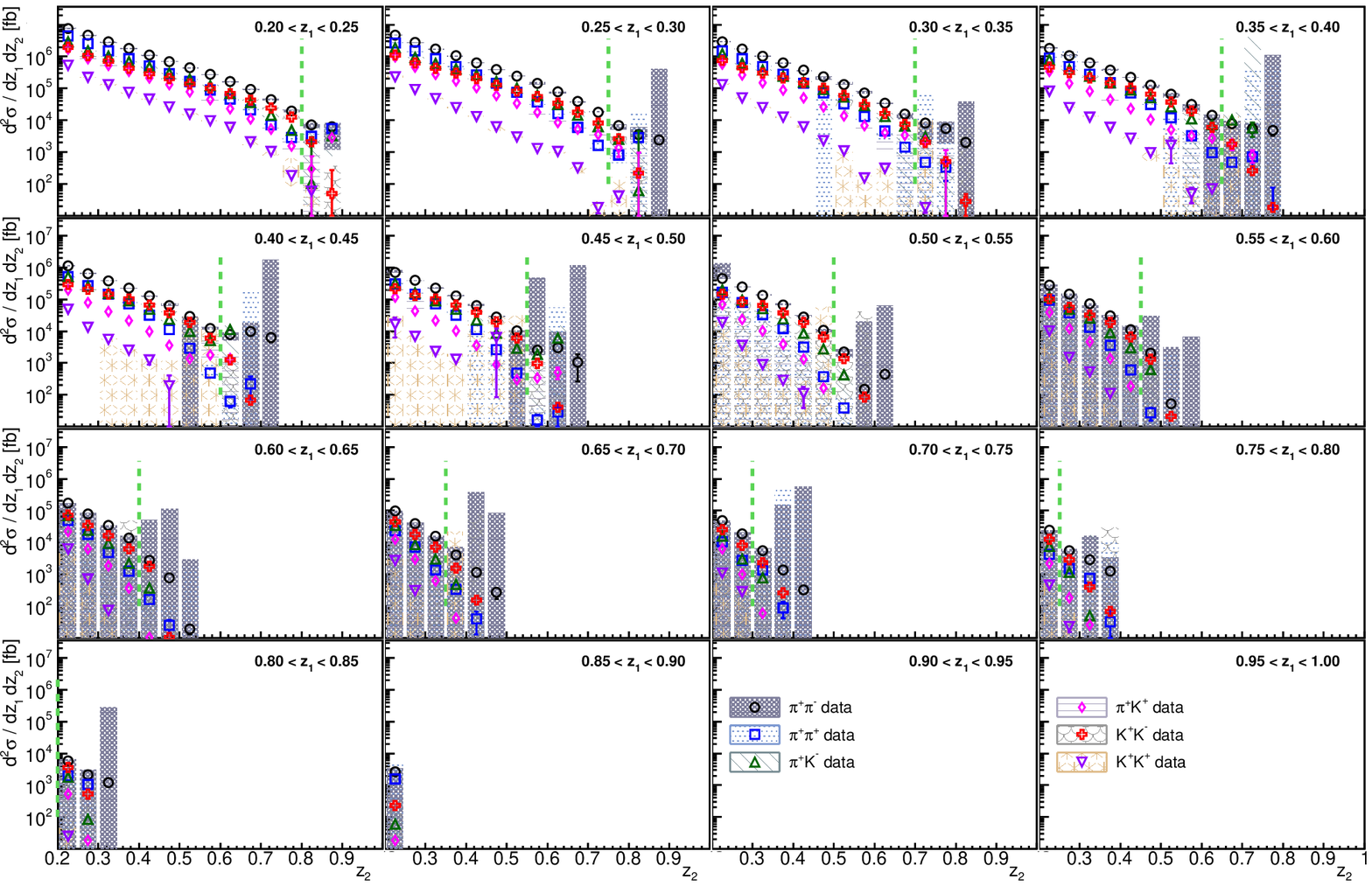}

\caption{\label{fig:yieldall_mix1_pid0_1_4_12}(color online) Differential cross sections for $\pi^+\pi^-$ (black circles), $\pi^+\pi^+$ (blue squares), $\pi^+K^-$ (green triangles), $\pi^+K^+$ (purple diamonds), $K^+K^-$ (red crosses) and $K^+K^+$ (violet downward triangles) pairs in the {\it same} topology (including $T>0.8$ selection) as a function of $z_2$ for the indicated $z_1$ bins. The error boxes represent the systematic uncertainties. The vertical, dashed green line represents the $z_1+z_2=1$ limit in each panel.}
\end{center}
\end{figure*}

The cross sections for di-hadrons in the {\it same} hemisphere are displayed in Fig.~\ref{fig:yieldall_mix1_pid0_1_4_12} for all ($z_1,z_2$) bins. The cross sections fall off rapidly and mostly disappear at the boundary $z_1+z_2 =1$, where the total energy of one initial parton is fully contained in the energy of the two hadrons. A small excess above this limit can be seen. MC studies show that this excess can be explained qualitatively by a small mis-assignment of hemisphere due to the smearing of the thrust axis relative to the initial quark-antiquark axis. In addition, hard gluon radiation may create such events. 

\subsubsection{Cross section ratios}
\begin{figure*}[htb]
\begin{center}
\includegraphics[width=0.9\textwidth]{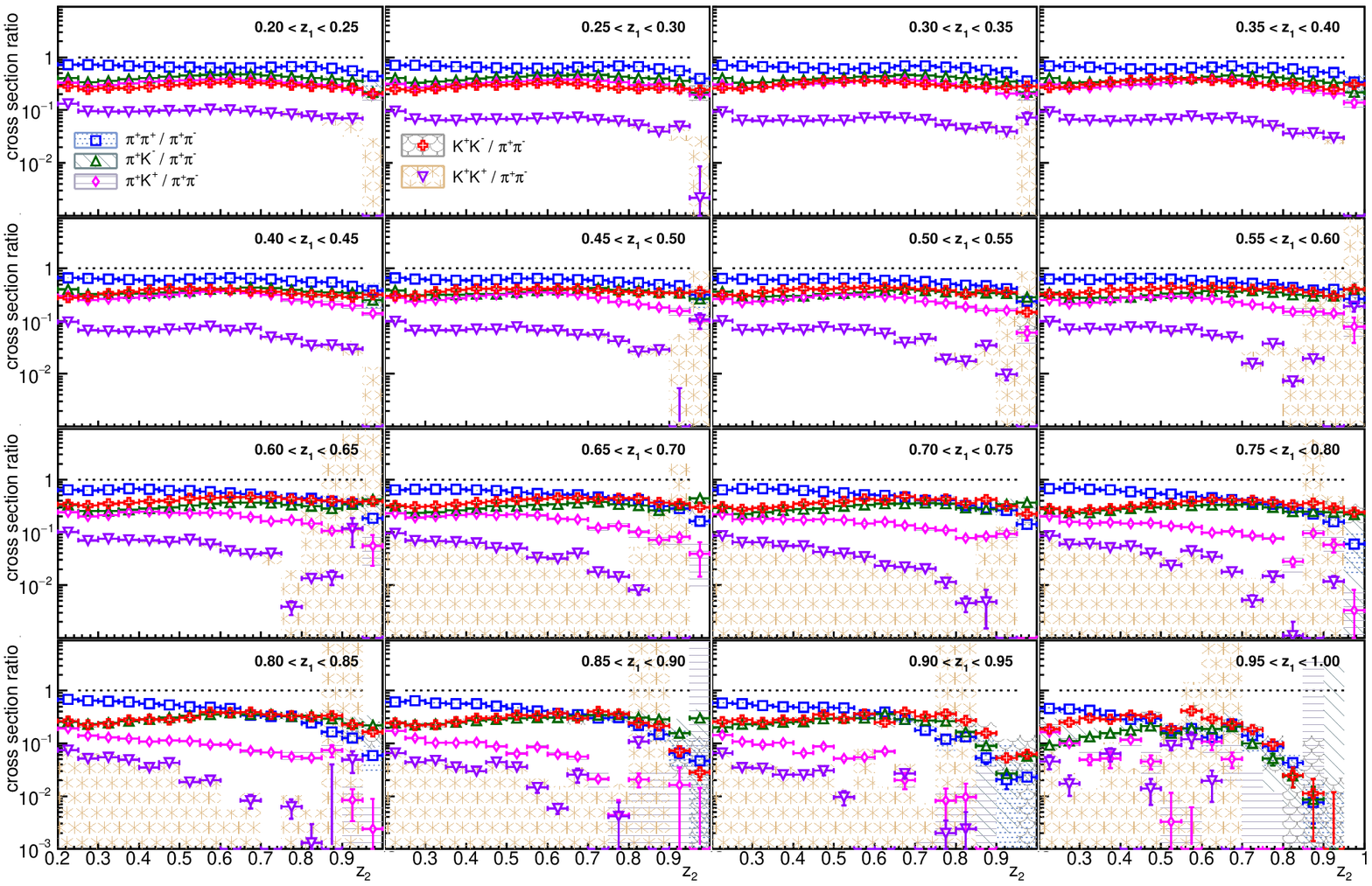}

\caption{\label{fig:allratio_pid15_mix16_c8}(color online) Ratios of differential cross sections for $\pi^+\pi^+$ (blue squares), $\pi^+K^-$ (green triangles), $\pi^+K^+$ (purple diamonds), $K^+K^-$ (red crosses) and $K^+K^+$ (violet downward triangles) pairs relative to the $\pi^+\pi^-$ pairs in the {\it any} topology as a function of $z_2$ for the indicated $z_1$ bins. The shaded areas correspond to the systematic uncertainties under the assumption that they are independent between each two di-hadron combinations.}
\end{center}
\end{figure*}
As several of the uncertainties are common to all charge and hadron combinations, these cancel in ratios and the information about favored and disfavored fragmentation should be more reliable. For example, ignoring strange-quark fragmentation to pions, the same-sign to opposite-sign pion-pair ratios are simple measures of disfavored \textit{vs.} favored pion fragmentation functions for light quarks. As can be seen in the ratios in Fig.~\ref{fig:allratio_pid15_mix16_c8}, they show a similar nearly flat behavior at low fractional energies, but deviate substantially from this trend at higher fractional energies.  
Generally, the opposite-sign pion-kaon and kaon-kaon ratios are more suppressed than the already disfavored same-sign pion pairs, but at the highest ($z_1,z_2$) do become comparable. This might be related to the fact that opposite-sign kaons can be favored fragmentation for both hadrons if a $s\bar{s}$ was created initially. As the opposite-sign pion-kaon pairs are similar in magnitude, even at high $z$, it can be argued that disfavored fragmentation from strange quarks to pions is not as suppressed. Same-sign kaon pairs, where at least one strange-quark pair needs to be produced in the fragmentation, are always suppressed at least one order of magnitude relative to the opposite-sign pion pairs. This shows that strangeness produced in fragmentation is indeed strongly suppressed, as is generally assumed in fragmentation models such as those included in {\sc Pythia}.

\subsubsection{Hemisphere decomposition}
\begin{figure*}[htb]
\begin{center}
\includegraphics[width=0.8\textwidth]{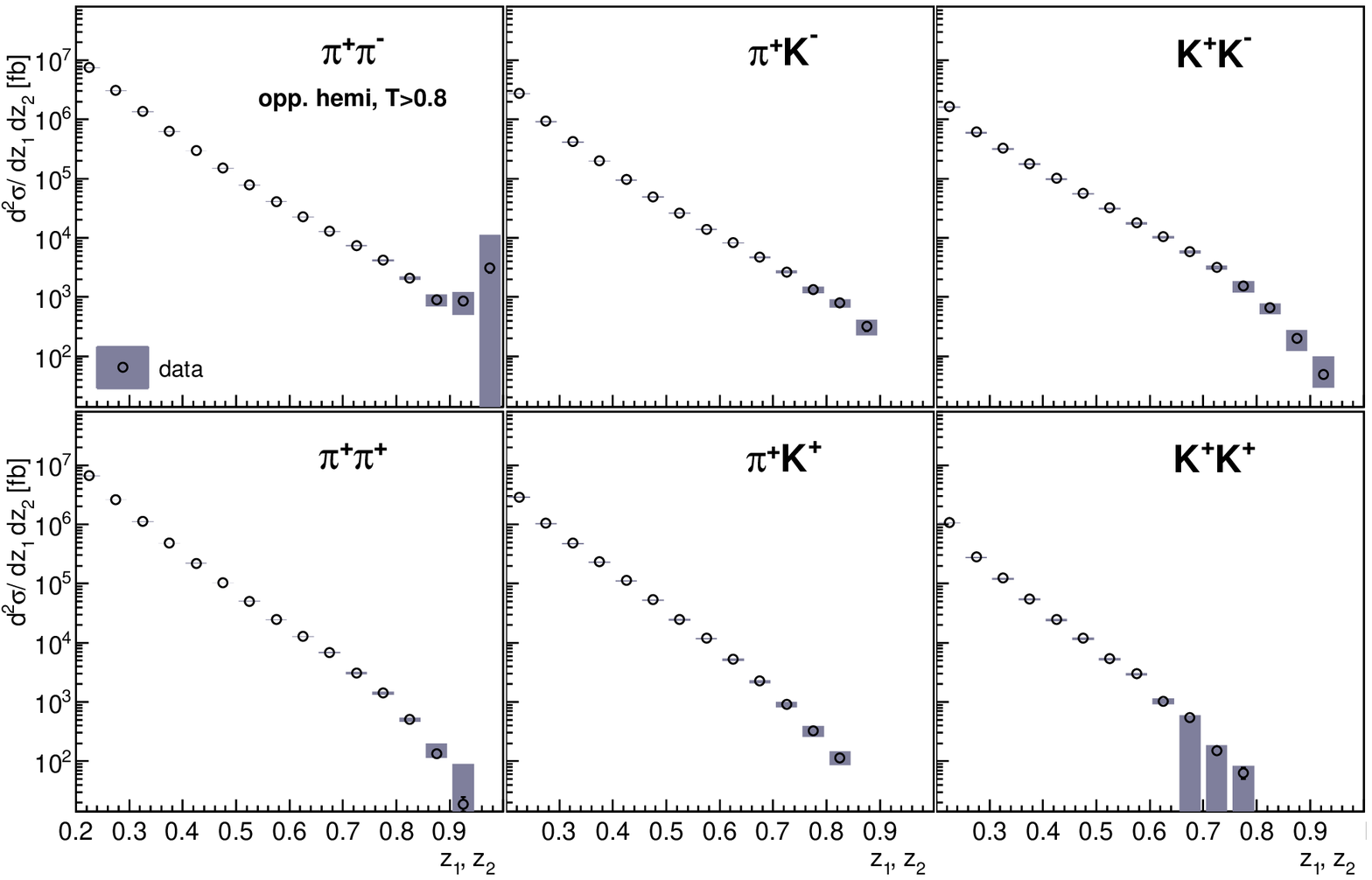}

\caption{\label{fig:xsecs} Differential cross sections for the main hadron pairs within {\it opposite} hemispheres (including thrust selection $T>0.8$) as a function of ($z_1,z_2$) for diagonal bins only. The error boxes represent the systematic uncertainties.}
\end{center}
\end{figure*}

\begin{figure*}[htb]
\begin{center}
\includegraphics[width=0.8\textwidth]{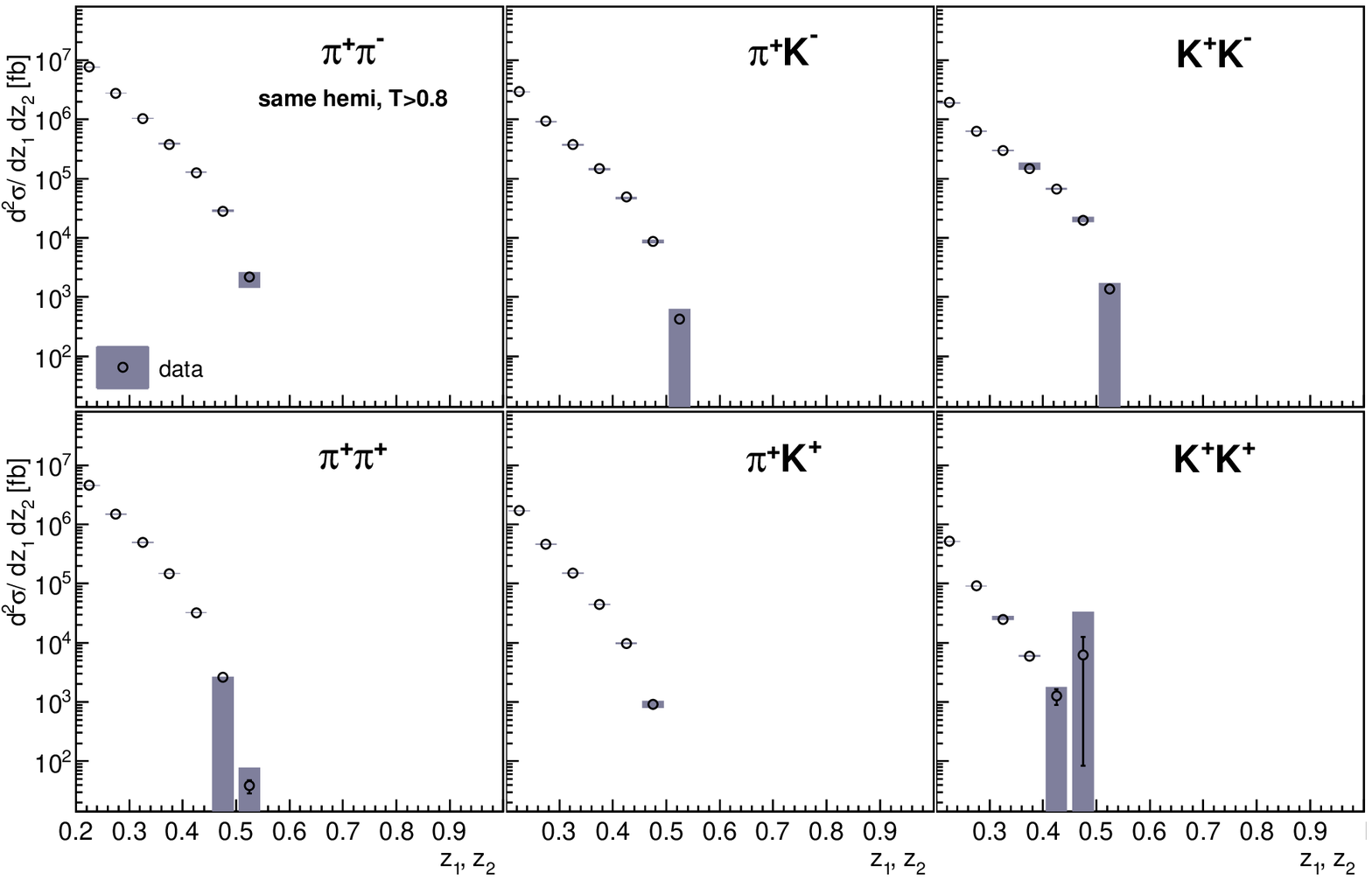}

\caption{\label{fig:xsecs2} Differential cross sections for the main hadron pairs within the {\it same} hemisphere (including thrust selection $T>0.8$) as a function of ($z_1,z_2$) for diagonal bins only. The error boxes represent the systematic uncertainties.}
\end{center}
\end{figure*}

Figure \ref{fig:xsecs} displays all six relevant hadron combinations for {\it opposite} hemispheres while Fig.~\ref{fig:xsecs2} shows the cross sections for hadrons within the {\it same} hemisphere using a thrust requirement of $T>0.8$. Note that the requirement of a minimum thrust value is not corrected for in these hemisphere decompositions, which must be taken into account when used for global FF analyses. As expected, the cross sections at small $z$ are of similar magnitude between the {\it same} and {\it opposite} hemispheres, while at higher $z$ only {\it opposite}-hemisphere pairs remain. 

\begin{figure*}[htb]
\begin{center}
\includegraphics[width=0.8\textwidth]{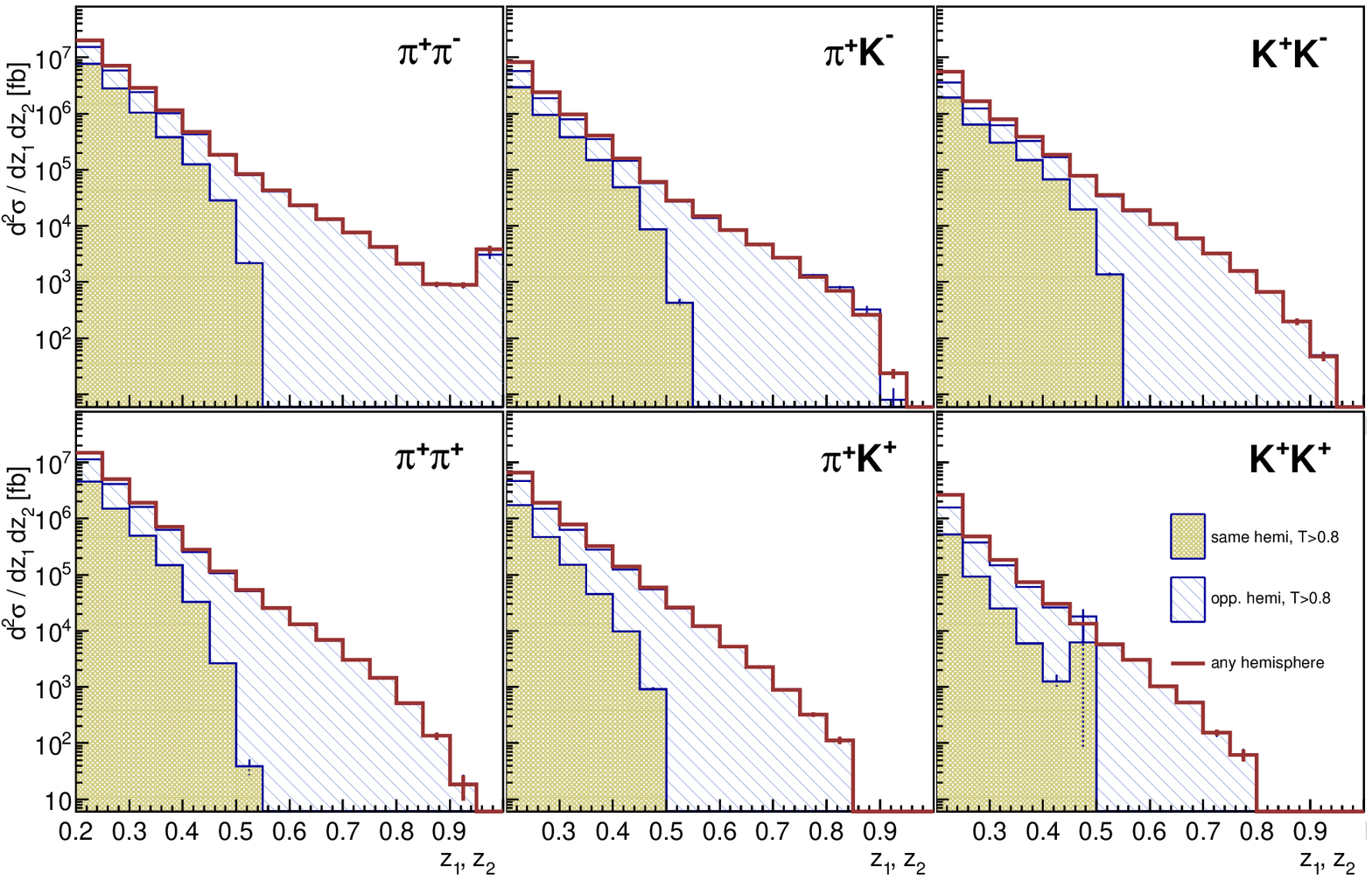}
\includegraphics[width=0.8\textwidth]{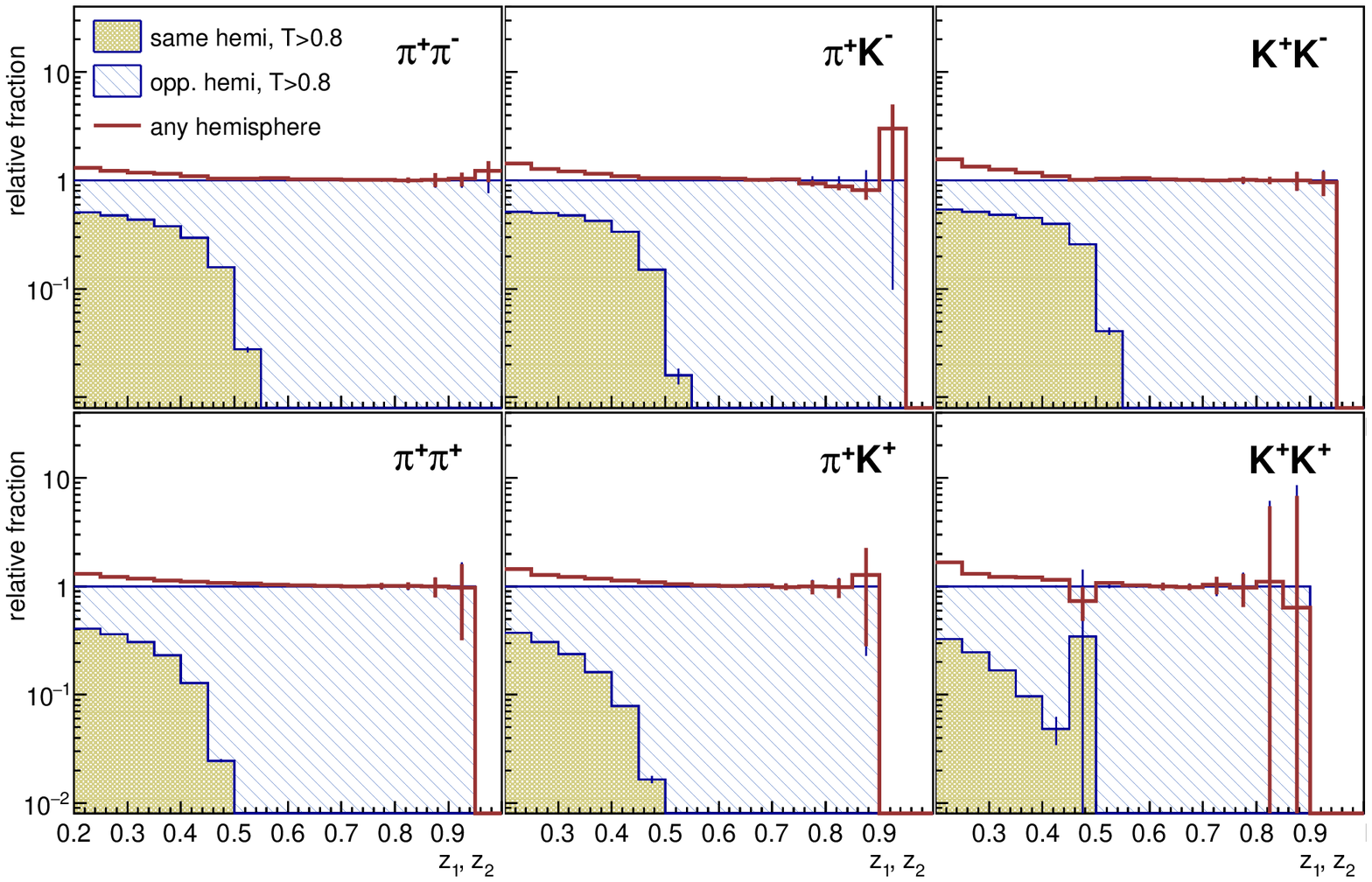}

\caption{\label{fig:plotmixhemisum_c8}(color online) Differential cross sections (top) and ratios to the {\it same} and {\it opposite} topology sum (bottom) for the main hadron combinations, stacking {\it same} (gray filled areas, including thrust selection T > 0.8) and {\it opposite} (blue hatched areas, including thrust selection T > 0.8) hemisphere data and comparing to those without hemisphere assignment (red curves). For visibility, only the diagonal ($z_1,z_2$) bins and statistical uncertainties are displayed.}
\end{center}
\end{figure*}

These cross sections with hemisphere assignment can be compared to the cross sections without the hemisphere assignment and without the thrust requirement, as shown in Fig.~\ref{fig:plotmixhemisum_c8} for diagonal ($z_1,z_2$) bins. As expected, the {\it same} hemisphere contributions are comparable to those from {\it opposite} hemispheres at very low $z$ while they vanish rapidly with increasing $z$. This is in agreement with the assumption of the {\it same} hemisphere di-hadrons emerging predominantly from the common initial parton. If so, the sum of the two hadron's fractional energies cannot exceed unity and they do indeed drop to zero at ($z_1,z_2$) each around 0.5. \par
The small difference between the sum of the {\it same} and {\it opposite} hemispheres and the {\it any} topology assignment at low $z$ is due to the additional thrust selection used to identify hemispheres. The small deviations seen occasionally at high $z$ are related to variations from the smearing unfolding and are consistent within the uncertainties that are not shown in this figure.

\subsubsection{MC generator comparison}
\begin{figure*}[htb]
\begin{center}
\includegraphics[width=0.8\textwidth]{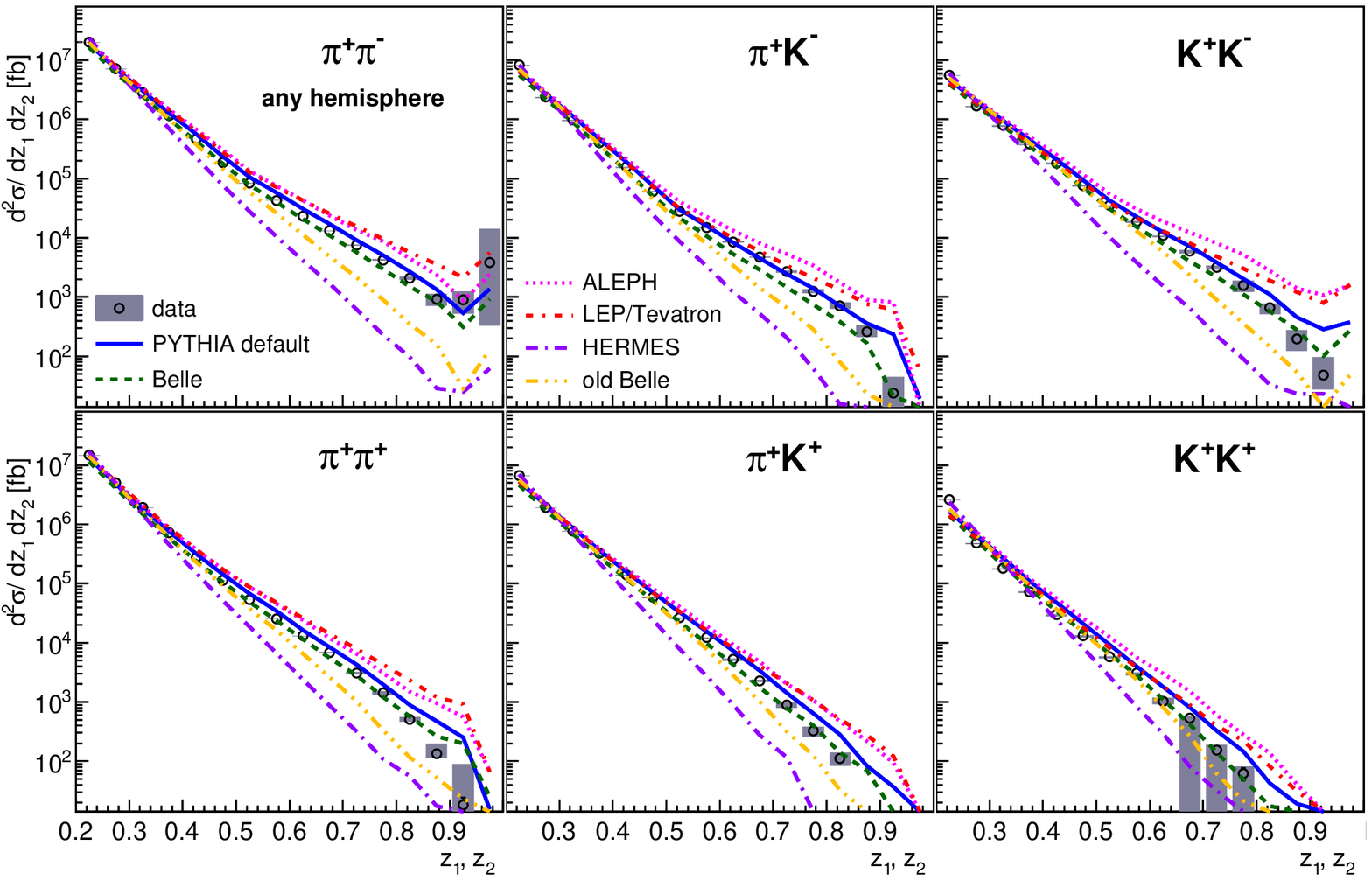}
\includegraphics[width=0.8\textwidth]{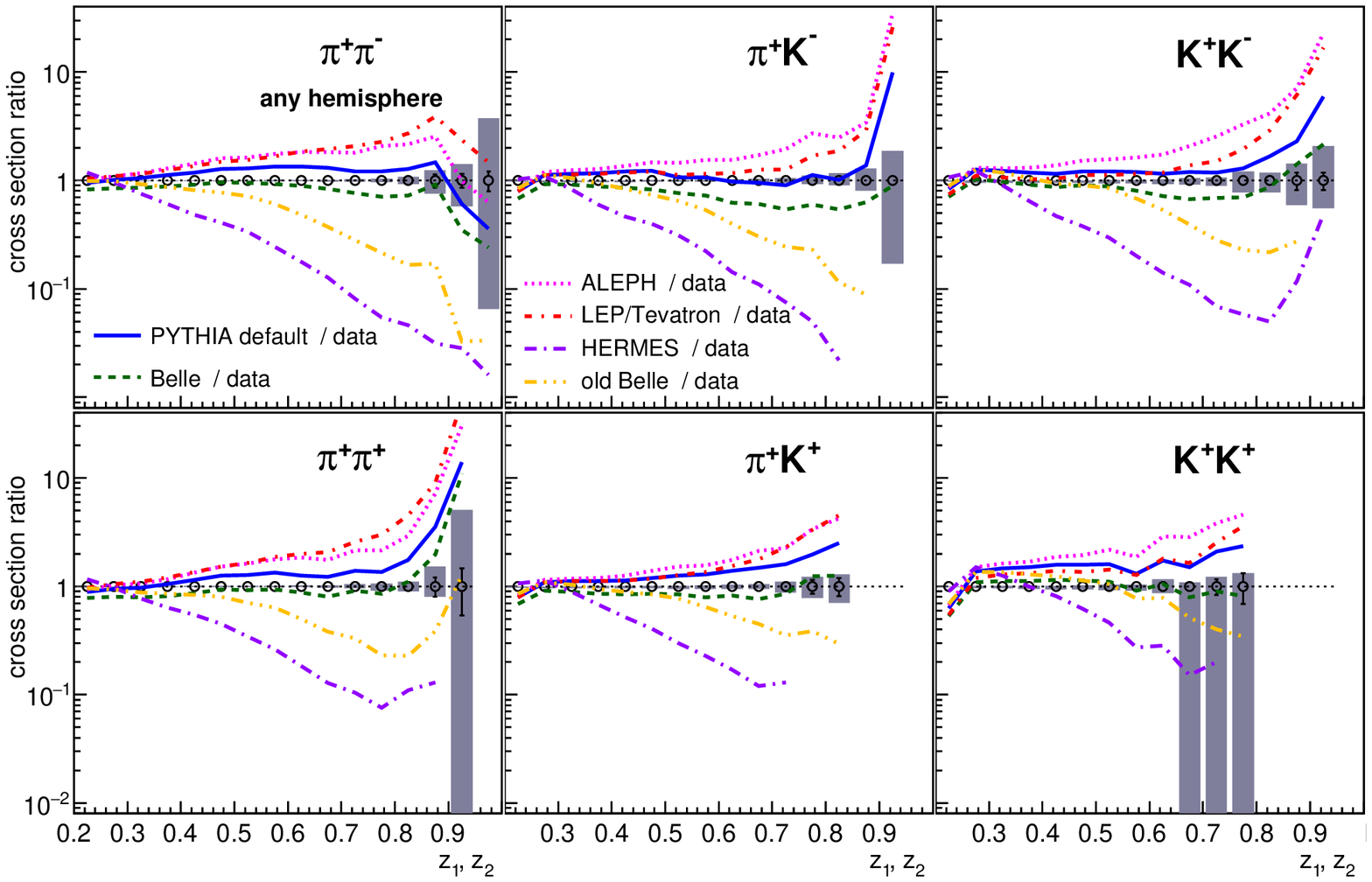}

\caption{\label{fig:allratiosum_mix16_c8_32518153}(color online) Differential cross sections (top) and ratios to the data (bottom) for the main hadron pairs in the {\it any} topology as a function of ($z_1,z_2$) for diagonal bins only. Various {\sc Pythia} tunes are also displayed as described in the text. For comparison, the relative statistical and systematic uncertainties are shown for the data as well.}
\end{center}
\end{figure*}

The di-hadron cross sections in the {\it any} topology are compared to various fragmentation settings within the {\sc Pythia}/JetSet MC generator. These are displayed in Fig.~\ref{fig:allratiosum_mix16_c8_32518153} for the main hadron combinations and  diagonal ($z_1,z_2$) bins. The settings correspond to the default {\sc Pythia}, those currently used in Belle for the fully tracked GEANT simulations, as well as various other settings tuned to specific experiments, collision systems, and energies from the LEP/Tevatron, ALEPH, and HERMES environments. Generally, all parameterizations agree at very low $z$ as the total production yields for certain particles are best known. However, at very high $z$, the distributions differ greatly. It appears that for all six hadron and charge combinations the default {\sc Pythia} and the latest Belle settings describe the data best even at large fractional energies. The LEP-based tunes generally overshoot the data at high $z$ while the older Belle and HERMES tunes fall off much too rapidly.   

In addition to the {\it any} topology combination, which is dominated by the {\it opposite} topology at higher fractional energies, the {\it same} hemisphere combination is compared to these MC tunes. In principle, the different hemisphere combinations are sensitive to different parameters in the {\sc Pythia} settings. An example of the comparison for $\pi^+K^-$ pairs is displayed in Fig.~\ref{fig:allxsec_pid16_mix1_c8_32518153}; other hadron combinations are available in the supplementary file. The overall behavior is similar to Fig.~\ref{fig:allratiosum_mix16_c8_32518153}, with the older Belle and HERMES tunes undershooting the data. The other parameterizations do not differ as substantially as in the {\it any} hemisphere combinations and they all reproduce the data reasonably well. 
The differences in the {\sc Pythia} settings are summarized in Table 1 of the supplementary file.

\begin{figure*}[htb]
\begin{center}
\includegraphics[width=0.9\textwidth]{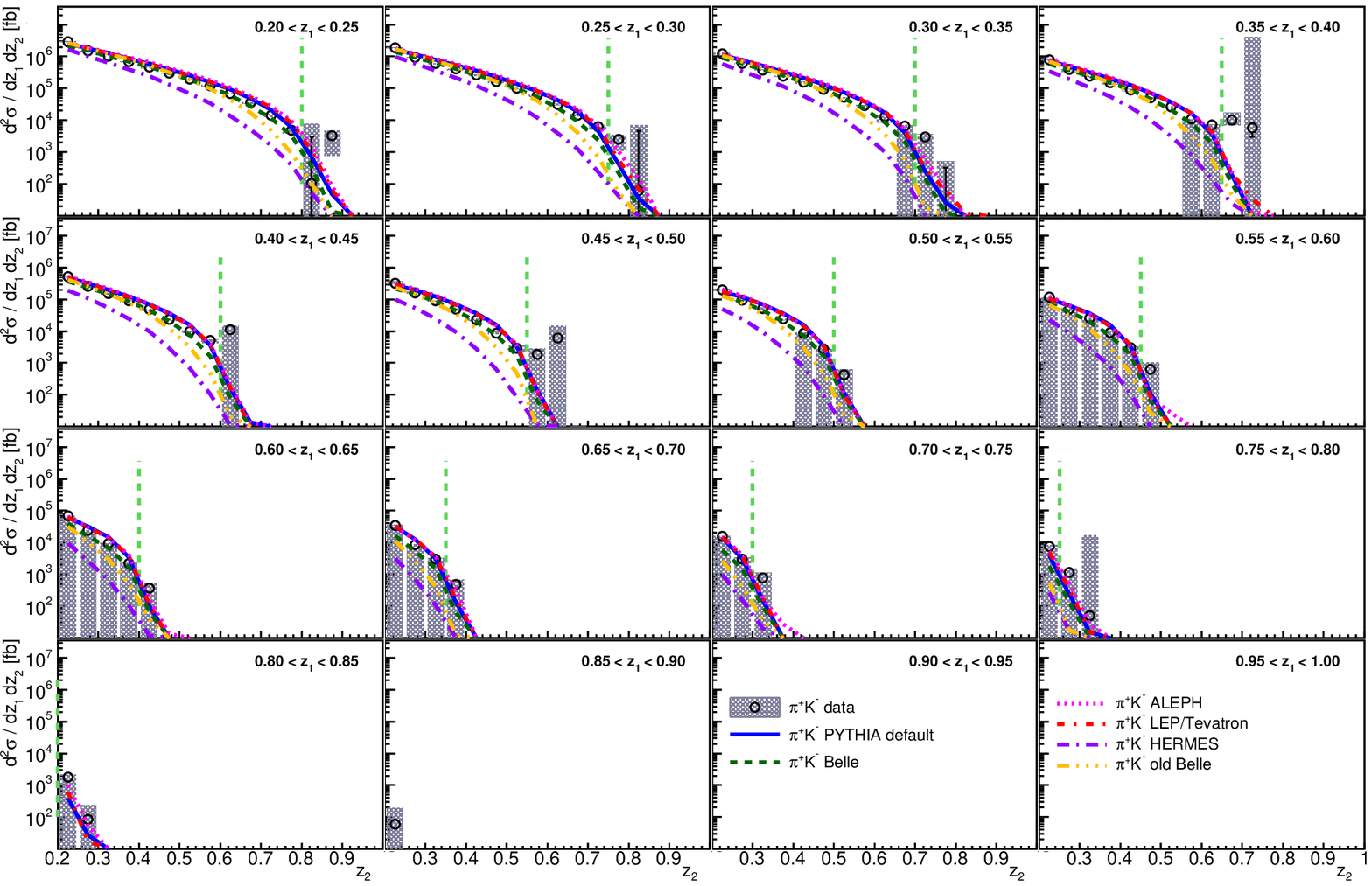}

\caption{\label{fig:allxsec_pid16_mix1_c8_32518153}(color online) Differential cross sections for $\pi^+K^-$ pairs in the {\it same} hemisphere as a function of ($z_1,z_2$). Various {\sc Pythia} tunes are also displayed as described in the text.}
\end{center}
\end{figure*}

\section{Single-hadron analysis}
\begin{figure*}[htb]
\begin{center}
\includegraphics[width=0.85\textwidth]{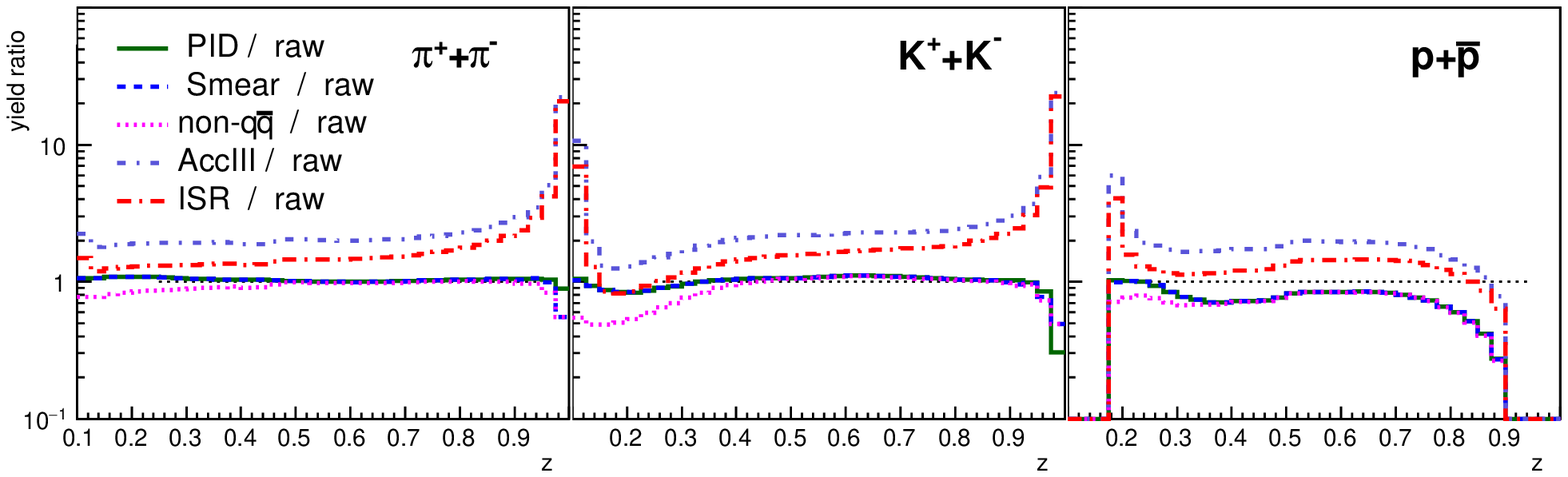}

\caption{\label{fig:singlebeforeafter}(color online) Ratio of yields after various corrections relative to the raw yields for the main single hadrons are shown as a function of $z$. From top to bottom, the ratios after acceptance correction (purple, dash-dotted line), initial state radiation (red, long dash-dotted line), PID correction (dark green, full lines), smearing correction (blue, dashed line) and non-$q\bar{q}$ removal (magenta, dotted lines) are shown. }
\end{center}
\end{figure*}

In addition to the di-hadron analysis, the production of single hadrons, especially previously unpublished single protons is considered here.
The single-hadron analysis follows the same procedure as the di-hadron analysis. The $z$ range between 0.1 and 1.0 is divided into 36 bins; for protons, $z<$ 0.2 is kinematically inaccessible. The particle misidentification correction is performed as in the di-hadron analysis (but only for one track) and the resulting yield modification is shown in Fig.~\ref{fig:singlebeforeafter}. At intermediate $z$, in particular, the proton yields are reduced substantially due to proton misidentification. Non-$q \bar{q}$ events contribute once again to the pion and kaon distributions but not as much to protons, where predominantly $eeu\bar{u}$ processes at high $z$ ($\approx 5\%$)  and $\Upsilon$ decays at low $z$ (maximally $\approx 20\%$) are the dominant backgrounds. All acceptance corrections are only weakly dependent on hadron type and show the same moderate (substantial) correction factors at small and intermediate (high) $z$; the high-$z$ correction is again dominated by the event preselection efficiencies. Weak decays originate predominantly from charm decays and so are a very small contribution ($< 10$\%) for protons. 
The various correction steps for single-pions, kaons and protons are summarized in Fig.~\ref{fig:singlebeforeafter}.
\begin{figure*}[t]
\begin{center}
\includegraphics[width=0.8\textwidth]{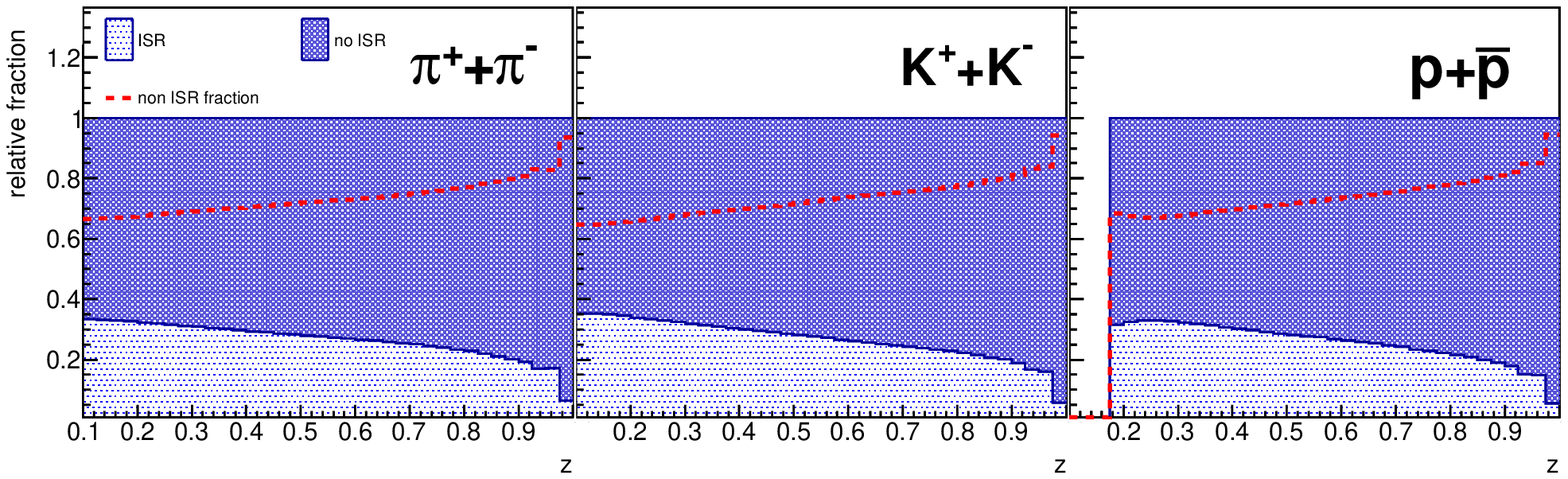}
\caption{\label{fig:singleisr}(color online) Relative fractions for pions, kaons and protons as a function of $z_1$ originating from ISR or non ISR events. The individual relative contributions are displayed from top to bottom for non ISR events (energy loss less than 0.5\%, purple, filled area) and ISR events (blue, hatched area) from generated generic $udsc$ MC. The relative fractions are also shown for the non-ISR fraction (green, dashed lines).}
\end{center}
\end{figure*}

The ISR correction here is similar to that in the di-hadron analysis. To clarify the correction for the previous single-pion and kaon results \cite{martin}, we show in Fig.~\ref{fig:singleisr} the ISR and non-ISR fractions for single pions and kaons as well as protons. As in the di-hadron analysis, the fraction of events with an actual CMS energy below 99.5\% of the nominal energy is below 30\% and decreases with increasing $z$. 
\begin{figure}[htb]
\begin{center}
\includegraphics[width=0.45\textwidth]{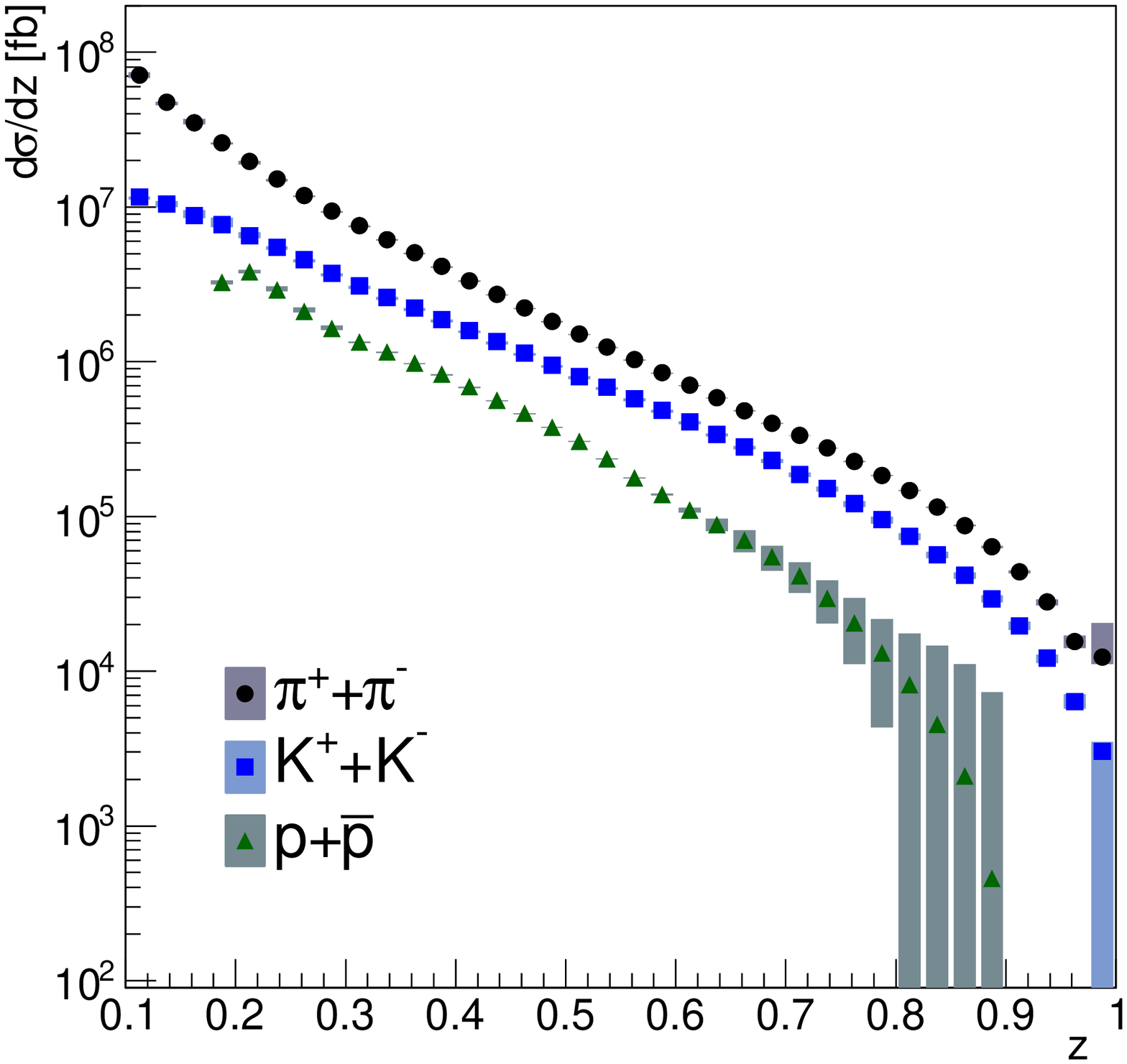}

\caption{\label{fig:singlehadronxsec}(color online) Single-pion (black circles), -kaon (blue squares) and -proton (green triangles) cross sections from top to bottom, as a function of $z$. }
\end{center}
\end{figure}

\begin{figure*}[htb]
\begin{center}
\includegraphics[width=0.85\textwidth]{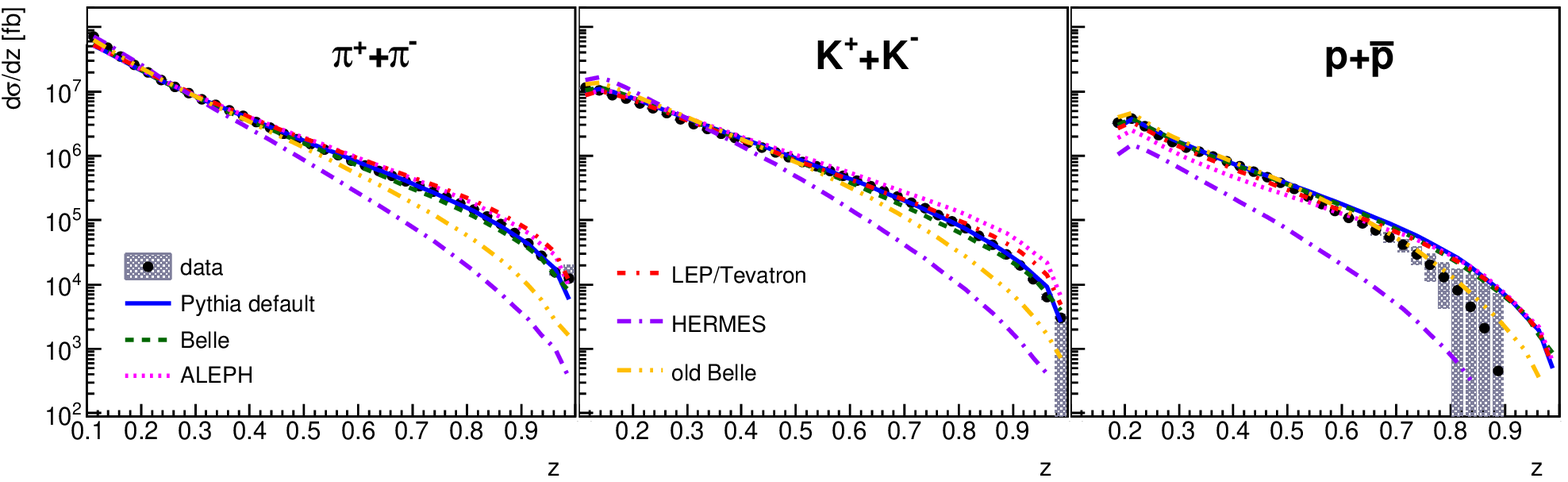}
\includegraphics[width=0.85\textwidth]{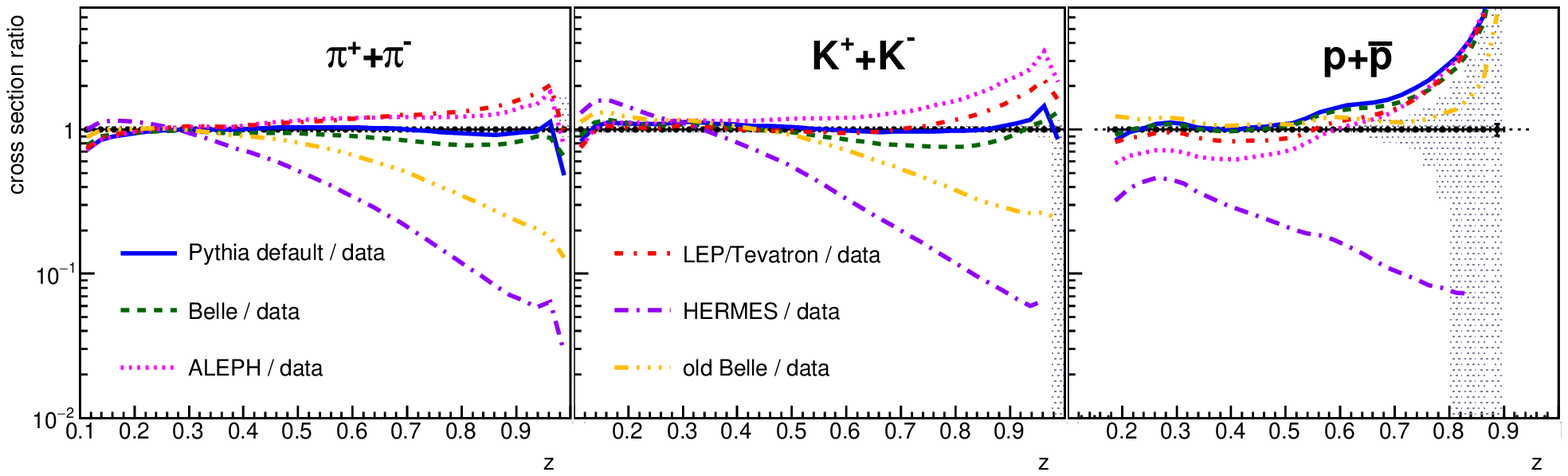}

\caption{\label{fig:singleallratio_mix4_c8_32518153}(color online) Differential cross sections (top) and ratios to the data (bottom) for the main single-hadrons as a function of $z$. Various MC tunes are also displayed as described in the text. For comparison, the relative statistical and systematic uncertainties are shown for the data as well.}
\end{center}
\end{figure*}

The resulting single-pion, -kaon, and -proton cross sections are displayed in Fig.~\ref{fig:singlehadronxsec}. While the pion and kaon results are consistent within uncertainties to those published before, the proton results from Belle are shown for the first time.
The results are compared with the aforementioned {\sc Pythia}/JetSet fragmentation tunes in Fig.~\ref{fig:singleallratio_mix4_c8_32518153}. As has been noted above and in \cite{martin}, the {\sc Pythia}/JetSet settings close to the default settings reproduce the pion and kaon cross sections rather well. For the proton cross sections, no setting describes the entire $z$ range, while the ALEPH and LEP/Tevatron tunes roughly agree with the data at low $z$ and the older Belle MC setting is in moderate agreement at high $z$. 

\section{Summary}
In summary, we present $e^+e^- \rightarrow h_1h_2X$ differential cross sections in $z_1$ and $z_2$ for pion-pion, pion-kaon and kaon-kaon pairs of the same and opposite charges and in various topologies. 
The general expectations of disfavored fragmentation functions being suppressed, especially at large fractional energies, are confirmed within the assumptions mentioned in this article. In particular, the same-sign pion pairs in {\it opposite} hemispheres fall off more rapidly than the opposite-sign pion pairs. 
The ordering with additional strangeness is also as expected when taking into account the favored-kaon fragmentation of strange quarks and charm decays. For example, where strangeness needs to be created in the fragmentation such as for same-sign kaon pairs and, to a lesser extent, the same-sign pion-kaon pairs, the cross sections decrease even more rapidly as the already disfavored same-sign pion pairs. 

The vanishing of the {\it same}-hemisphere di-hadron cross sections once the sum of the fractional energies of the two hadrons exceeds unity supports the assumption of the {\it same} hemisphere di-hadrons being produced predominantly via single-parton di-hadron fragmentation. This, in turn, bolsters the interpretation of the {\it opposite} hemisphere hadron pairs as arising from the fragmentation of different partons. As a consequence, the inclusion of the {\it opposite} hemisphere di-hadrons in terms of single-hadron fragmentation into a global pQCD fragmentation function analysis should be possible. 

The extracted di-hadron cross sections are compared to various {\sc Pythia} MC tunes that were optimized for various other energies and collision systems. A full optimization at Belle energies should be possible based on these results; nevertheless, both the default {\sc Pythia} fragmentation setting as well as the latest Belle fragmentation setting already describe the data reasonably well.
 
Single-proton $e^+e^- \rightarrow pX$ cross sections differential in $z$ are presented in addition to the previously published single-pion and -kaon results. These are expected to be of use in global analyses of fragmentation functions, including the proton results from BaBar \cite{babar}. Various {\sc Pythia} tunes are compared with conclusions similar to those in the di-hadron case. However, for the proton cross sections, the agreement is fair at best over the entire $z$ range, which suggests room for improvement in the {\sc Pythia} settings to better model the baryon production.

With the precision of these measurements and the additional information obtained by the use of di-hadrons, we expect that subsequent global fits to the world data will improve substantially our understanding of fragmentation functions, in particular in terms of the distinction of favored versus disfavored fragmentation.  
\begin{acknowledgments}
We thank the KEKB group for the excellent operation of the
accelerator; the KEK cryogenics group for the efficient
operation of the solenoid; and the KEK computer group,
the National Institute of Informatics, and the 
PNNL/EMSL computing group for valuable computing
and SINET4 network support.  We acknowledge support from
the Ministry of Education, Culture, Sports, Science, and
Technology (MEXT) of Japan, the Japan Society for the 
Promotion of Science (JSPS), and the Tau-Lepton Physics 
Research Center of Nagoya University; 
the Australian Research Council and the Australian 
Department of Industry, Innovation, Science and Research;
Austrian Science Fund under Grant No.~P 22742-N16 and P 26794-N20;
the National Natural Science Foundation of China under Contracts 
No.~10575109, No.~10775142, No.~10875115, No.~11175187, and  No.~11475187; 
the Chinese Academy of Science Center for Excellence in Particle Physics; 
the Ministry of Education, Youth and Sports of the Czech
Republic under Contract No.~LG14034;
the Carl Zeiss Foundation, the Deutsche Forschungsgemeinschaft
and the VolkswagenStiftung;
the Department of Science and Technology of India; 
the Istituto Nazionale di Fisica Nucleare of Italy; 
National Research Foundation (NRF) of Korea Grants
No.~2011-0029457, No.~2012-0008143, No.~2012R1A1A2008330, 
No.~2013R1A1A3007772, No.~2014R1A2A2A01005286, No.~2014R1A2A2A01002734, 
No.~2014R1A1A2006456;
the Basic Research Lab program under NRF Grant No.~KRF-2011-0020333, 
No.~KRF-2011-0021196, Center for Korean J-PARC Users, No.~NRF-2013K1A3A7A06056592; 
the Brain Korea 21-Plus program and the Global Science Experimental Data 
Hub Center of the Korea Institute of Science and Technology Information;
the Polish Ministry of Science and Higher Education and 
the National Science Center;
the Ministry of Education and Science of the Russian Federation and
the Russian Foundation for Basic Research;
the Slovenian Research Agency;
the Basque Foundation for Science (IKERBASQUE) and 
the Euskal Herriko Unibertsitatea (UPV/EHU) under program UFI 11/55 (Spain);
the Swiss National Science Foundation; the National Science Council
and the Ministry of Education of Taiwan; and the U.S.\
Department of Energy and the National Science Foundation.
This work is supported by a Grant-in-Aid from MEXT for 
Science Research in a Priority Area (``New Development of 
Flavor Physics'') and from JSPS for Creative Scientific 
Research (``Evolution of Tau-lepton Physics'').
\end{acknowledgments}

\end{document}